\documentclass[12pt]{iopart}

\usepackage{amssymb}
\usepackage{graphicx,tabularx}
\usepackage{citesort}

\newcommand{\wt}{\widetilde}

\begin{document}


\title
[Long-lived charged particles produced by atmospheric and astrophysical
 neutrinos]
{Probing new physics with long-lived charged particles produced by 
 atmospheric and astrophysical neutrinos}

\author{Shin'ichiro Ando$^{1}$, John F. Beacom$^{2,3,4}$, Stefano
Profumo$^{5,1}$, and David Rainwater$^6$}

\vspace*{0.6cm}

\address{$^1$ California Institute of Technology, Pasadena, CA 91125,
USA}
\address{$^2$ Department of Physics, The Ohio State University, Columbus,
OH 43210, USA}
\address{$^3$ Department of Astronomy, The Ohio State University,
Columbus, OH 43210, USA}
\address{$^4$ Center for Cosmology and Astro-Particle Physics, The Ohio
State University, Columbus, OH 43210, USA}
\address{$^5$ Santa Cruz Institute for Particle Physics, University of
California, Santa Cruz, CA 95064, USA}
\address{$^6$ Department of Physics and Astronomy, University of
Rochester, Rochester, NY 14627, USA}

\eads{\\
\mailto{ando@tapir.caltech.edu},
\mailto{beacom@mps.ohio-state.edu},
\mailto{profumo@caltech.edu},
\mailto{rain@pas.rochester.edu}}

\vspace*{0.6cm}

\begin{abstract}
As suggested by some extensions of the Standard Model of particle
physics, dark matter may be a super-weakly interacting lightest stable
particle, while the next-to-lightest particle (NLP) is charged and
meta-stable.  One could test such a possibility with neutrino
telescopes, by detecting the charged NLPs produced in high-energy
neutrino collisions with Earth matter.  We study the production of
charged NLPs by both atmospheric and astrophysical neutrinos; only the
latter, which is largely uncertain and has not been detected yet, was
the focus of previous studies.  We compute the resulting fluxes of the
charged NLPs, compare those of different origins, and analyze the
dependence on the underlying particle physics setup.  We point out
that even if the astrophysical neutrino flux is very small,
atmospheric neutrinos, especially those from the prompt decay of
charmed mesons, may provide a detectable flux of NLP pairs at neutrino
telescopes such as IceCube.  We also comment on the flux of charged
NLPs expected from proton--nucleon collisions, and show that, for
theoretically motivated and phenomenologically viable models, it is
typically sub-dominant and below detectable rates.
\end{abstract}

\maketitle


\section{Introduction}
\label{sec:intro}

The fundamental nature of dark matter poses a profound challenge
to contemporary theoretical particle physics. Observations constrain
the neutrino---the only electrically and color neutral non-baryonic
elementary particle within the Standard Model of particle physics---to
have a negligible contribution to the overall dark matter 
budget~\cite{Fukugita:2006rm,Spergel:2006hy}. Dark matter is regarded
as one of the most compelling hints towards new physics beyond the
Standard Model. The question of its elementary essence has triggered
enormous theoretical and phenomenological efforts~\cite{Bergstrom:2000pn}.

The existence of suitable dark matter particle candidates in several
theoretically cogent extensions of the Standard Model, like low-energy
supersymmetry~\cite{Jungman:1995df} or extra-dimensional
scenarios~\cite{Cheng:2002ej,Hooper:2007qk}, focused a strong interest
on a class of dark matter candidates known as weakly interacting
massive particles (WIMPs). Similarly to other Standard Model
particles, WIMPs would fall out of thermal equilibrium and freeze out
in the early Universe, leaving a relic abundance compatible with the
inferred amount of dark matter~\cite{Wolfram:1978gp}.
These WIMPs can be directly detected by experiments looking for the
minuscule energy deposition caused by dark matter particles scattering
nuclei~\cite{Munoz:2003gx}. The pair annihilation
of WIMPs into energetic gamma rays, neutrinos, and antimatter, is a
second, yet indirect, handle on the presence and potential imprint of
galactic particle dark matter~\cite{Bergstrom:1998hd}.

The connection of the aforementioned scenarios to the electroweak
scale, soon to be probed at the Large Hadron Collider (LHC), motivated
the exploration of complementarity between collider searches for new
physics and the question of the elementary nature of dark
matter~\cite{Drees:2000he,Baer:2003wx,Weiglein:2004hn,Masiero:2004ft,Baltz:2006fm}.  
In most cases, if dark matter is a WIMP,
the anticipated experimental signature at LHC would be the production
of strongly interacting massive particles which promptly decay to the
lightest and stable WIMP, plus a number of energetic jets and leptons.
The neutral dark matter particle would escape the detector unobserved,
leading to large missing transverse energy as well~\cite{howie}.
Conclusive identification of escaping neutral particles at LHC with
dark matter permeating our Galaxy and other cosmic structures would,
however, require some evidence from the direct and/or indirect WIMP
searches listed before~\cite{Bergstrom:1998hd}.

WIMPs are indeed attractive dark matter candidates, but are not the
only theoretically envisioned possibility.  The dark matter particle
could exhibit even feebler interactions with ordinary Standard Model
particles than a WIMP, making direct and indirect searches completely
hopeless.  For instance, the supersymmetric
gravitino~\cite{Blumenthal:1982mv} or the Kaluza-Klein graviton of
universal extra dimensions~\cite{Feng:2003nr,Hooper:2007qk} are
perfectly plausible `super-weakly interacting'~\cite{Feng:2003xh} dark
matter candidates (super-WIMPs).  If Nature chose an option like this,
collider signatures of new physics, if any, would strongly depend upon
the nature of the next-to-lightest supersymmetric particle (NLSP).
Since a super-WIMP is also very weakly coupled to the other
new-physics heavier states, the NLSP would likely be quasi-stable.  If
the NLSP is neutral, the qualitative experimental landscape would look
like that of a standard WIMP scenario.  However, we would lack the
needed proof of a connection between the weakly interacting long-lived
particles produced at colliders and galactic dark matter.

If the NLSP is instead charged (constituting a charged massive
particle, or CHAMP), the LHC would potentially observe the extremely
distinct signature of a `heavy muon': charged tracks and penetration
of the outer muon sub-detector, possibly at very low relativistic
beta.  CHAMPs are constrained by direct collider searches at
LEP2~\cite{lep2}, as well as at the
Tevatron~\cite{tevatron}.\footnote{These and future searches at LHC
are not trivial, but advanced work is being done to ensure that such
potential signals would not be missed~\cite{Fairbairn:2006gg}.
Detection for masses $\lesssim 1$~TeV is essentially guaranteed.}

If a CHAMP were stable on collider scales, it could still decay on
cosmological scales, and thus impact precision astrophysical
measurements~\cite{chargeddecay}, including the chemical potential
associated with the cosmic microwave background black-body
spectrum~\cite{Zhang:2007zzh}, the extragalactic gamma-ray
background~\cite{Cembranos:2007fj}, the reionization history of the
universe~\cite{Pierpaoli:2003rz}, the formation of small scale
structures~\cite{Sigurdson:2003vy,Profumo:2004qt} and the synthesis of
light elements in the early Universe~\cite{BBNLDP,Jedamzik:2007cp}
(see, for implications of neutral particle, Ref.~\cite{BBNLDP2}).
Anomalies in the above mentioned quantities, however, could hardly be
considered smoking gun evidence that the collider CHAMPs were indeed
related to lighter, super-weakly interacting dark matter.

If CHAMPs were stable on collider timescales, but featured a short
lifetime on cosmological timescales, say on the order of a year or
less, CHAMPs produced in colliders might be trapped in large water
tanks surrounding the detectors~\cite{Feng:2004yi}.  The tanks would
then be periodically drained to underground reservoirs where CHAMP
decays might be observed in low-background
conditions~\cite{Feng:2004yi}.  While certainly not straightforward
experimentally, such a technique might provide important information
first on the actual meta-stability of the charged species, and,
secondly on the nature of the super-weakly-interacting particle the
CHAMP would decay into.  However, even if CHAMP decays were actually
observed, this would still not suffice as conclusive evidence that the
elusive particle CHAMPs decay into is indeed the dark matter
constituent.

To our knowledge, beyond high-energy collider experiments, the only
direct experimental handle on a super-weakly interacting dark matter
particle featuring a heavier, meta-stable charged partner is CHAMP
pair production via neutrino--nucleon collisions, followed by direct
observation at neutrino telescopes.  This idea, originally proposed in
Ref.~\cite{Albuquerque:2003mi}, relies on the fact that the energy
losses of CHAMPs in Earth are significantly smaller than those of
muons, therefore CHAMP pairs (unlike muon pairs) can reach the
detector even if they were produced far away.  This makes the relevant
target volume for neutrino--nucleon interactions much larger.
The CHAMP pairs can be efficiently separated from muon pairs, due
to large track separations in the detector.
The original proposal was subsequently followed up by related
studies~\cite{Ahlers:2006pf,Albuquerque:2006am,Reno:2005si,Huang:2006ie,Ahlers:2007js,Reno:2007kz},
which focused on the specific case of a gravitino lightest
supersymmetric particle (LSP), and a stau NLSP playing the role of the
CHAMP.  Among other aspects, these studies investigated in detail stau
energy losses in Earth and in the detector, computed expected
event rates for a few sample models and the relevant background, and
addressed the possibility of discriminating single-stau from
single-muon events.

Given the steady progress in the deployment of next-generation
km${}^3$-size neutrino telescopes---particularly IceCube at the South
Pole, already under construction and
taking data---we consider it timely to address in detail
a few points relevant for improving our understanding of the prospects
for implementing the above outlined technique. In particular, as
background rejection is not a substantial issue, the crucial point
appears to be the evaluation of the CHAMP pair event rate at IceCube.
To this end, we focus on the following four aspects:
\begin{itemize}
\item[(i)]

{\em Incoming neutrino flux}: So far, all long-lived CHAMP analyses for
neutrino telescopes have considered a relatively optimistic flux of
astrophysical neutrinos---as large as the Waxman-Bahcall (WB)
bound~\cite{WBBound}---as the primary source for CHAMP pair
production.  However, the WB bound is only a {\it theoretical upper
limit} on the flux of astrophysical neutrinos expected from optically
thin sources; therefore, the absolute normalization as well as the
spectral shape of the true astrophysical neutrino flux remain largely
unknown (of course the biggest reason for this is that these neutrinos
are as yet undetected!).  On the other hand, atmospheric neutrinos
have been detected and their flux is rather accurately known (below
$\sim$100~TeV~\cite{AtmExp}).  At larger energies ($\gtrsim 100$~TeV),
although there is no detection so far, one can rather reliably
extrapolate the flux of conventional atmospheric neutrinos from lower
energies.  In addition, one also expects a significant flux of
so-called prompt-decay atmospheric neutrinos, which originate from the
decay of short-lived charmed mesons and feature a harder spectral
index than the conventional component.  While we again have only an
upper limit on this prompt decay component, we know from particle
physics that it necessarily exists, at some level, in the high-energy
regime, and will be measured accurately by IceCube.  In any event,
atmospheric conventional and prompt-decay neutrinos evidently
contribute as well to CHAMP pair production in neutrino--nucleon
collisions.  In this paper, we study the role of these standard, {\it
guaranteed} neutrino sources, and compare it to the contribution from
astrophysical neutrino flux models including the WB bound.

\item[(ii)]

{\em Underlying particle physics model}: The event rate depends not
only on the incoming neutrino flux, but also on the nature of the
assumed particle physics model.  Here, we consider generic
supersymmetric models featuring a gravitino LSP and stau NLSP, and
study how the stau pair production cross section and event rates at
neutrino telescopes depend on the given mass spectrum.
Neutrino--nucleon interactions produce slepton--squark pair final
states, the amplitudes mediated by supersymmetric fermion exchange
(neutral and charged gauginos and higgsinos).

\item[(iii)] {\em Other CHAMP sources}: Unlike the high-energy
neutrino flux, the flux of very energetic protons is accurately
measured and well-known up to extremely high energies; proton--nucleon
interactions feature a supersymmetric pair production cross section
significantly larger than that of neutrino--nucleon processes. It
therefore seems reasonable to quantitatively assess the flux of stau
pairs produced in the interaction of incident primary protons with
nuclei in the atmosphere. The trade-off is the enormous cross section
for proton--nucleus scattering into Standard Model particles, which depletes the
incoming proton flux, highly suppressing any stau pair production
rate. Since quark--anti-quark processes can directly produce stau
pairs, however, the kinematic threshold for stau pair production, as a
function of the incoming primary particle energy, is lower for
proton--nucleus than for neutrino--nucleus processes. Yet we find that,
for reasonable and phenomenologically acceptable particle models, the
expected stau pair event rates from proton--nucleus collisions are
experimentally negligible at all energies, leaving neutrinos as the
only relevant primary source particles for CHAMP pair production.

\item[(iv)] {\em Simplified analytic approach:} We present the
computation of the flux of staus from neutrino--nucleon interactions
from first principles, and argue that, to an acceptable degree of
accuracy, the total number of expected staus can be computed as one
simple integral of three factors. Specifically, we show that the
quantities of physical relevance are (1) the incident flux of primary
neutrinos, (2) the ratio of the cross sections of neutrino plus nucleon
into supersymmetric particle pairs to the total neutrino--nucleon cross
section, and (3) a geometric efficiency factor.

\end{itemize}

Hereafter, we specifically use supersymmetric staus as charged
meta-stable NLPs, but note that the following arguments are applicable
to any other possible candidates of long-lived CHAMPs.
The outline of the remainder of the paper is as follows.  We discuss
the various components of the high energy neutrino flux and their
connected uncertainties in \Sref{sec:nuflux}.  We introduce the
new physics scenarios and compute the stau pair production cross
sections in \Sref{sec:particle}. \Sref{sec:rates} outlines the 
computation of the stau event rate at IceCube, including the
above-mentioned simplified analytic treatment.  The stau rate
dependence on the particle physics framework is addressed in 
\Sref{sec:susy}, and we draw conclusions in \Sref{sec:concl}.

\section{High-energy neutrino flux}\label{sec:nuflux}

In this Section, we summarize the high-energy neutrino fluxes we
consider in the present study. As discussed in \Sref{sec:intro}, past
works considered only a flux of neutrinos close to, or saturating, the
WB upper limit. However, other neutrino sources potentially contribute
as well: these include conventional atmospheric neutrinos,
prompt-decay atmospheric neutrinos and possibly astrophysical
neutrinos other than those considered in the WB setup.
Since the stau production rate does not depend on neutrino
flavor,\footnote{Flavor is conserved in the underlying supersymmetric
particle pair production event, but all supersymmetry particles then decay
promptly to a stau plus Standard Model particles.  Thus, an electron
neutrino eventually produces a stau pair just as a tau neutrino does.}
all we care is the total neutrino plus anti-neutrino flux, {\em i.e.},
the flux of $\nu=\nu_e+\nu_\mu+\nu_\tau$, where each $\nu_i$ here
indicates neutrino {\it plus} anti-neutrino of flavor $i$.  From this
point on, we mean this combined quantity when we use the term `flux',
unless otherwise stated. In \Fref{flux}, we summarize and collect the
various neutrino sources we discuss below, and the ranges of
normalizations we consider.

\subsection{High-energy astrophysical neutrinos from extra-galactic
  sources}

Very powerful astrophysical objects such as active galactic nuclei
(AGNs) and gamma-ray bursts (GRBs) are candidate high-energy neutrino
sources.
This is because strong gamma-ray emission detected from
these objects can be attributed to the particle acceleration and
successive interaction with the surrounding medium, magnetic and
photon fields, which might also be a source of neutrinos via charged
meson production.

\begin{figure}[t]
\begin{center}
\includegraphics[width=11cm,clip]{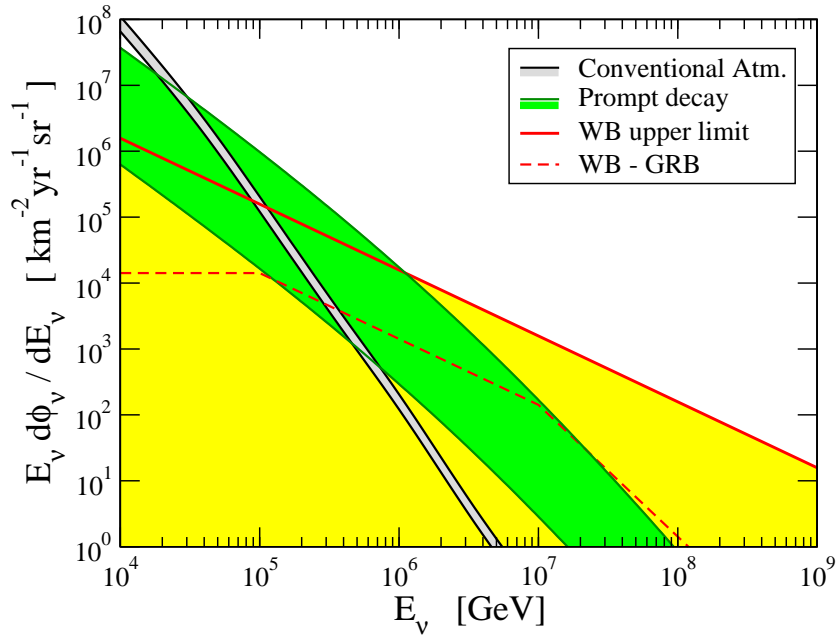}
\caption{The differential flux of high-energy neutrinos (plus
anti-neutrinos) considered in the present study, summed over neutrino
flavors.  The shaded gray, green and yellow regions indicate the
uncertainty ranges for the fluxes of conventional atmospheric,
prompt-decay atmospheric, and astrophysical neutrinos, respectively.
The reference extra-galactic neutrino fluxes refer to the WB
limit~\protect\cite{WBBound} (red solid line) and to their prediction
for GRBs~\cite{WBGRB} (red dashed line).}
\label{flux}
\vspace{-5mm}
\end{center}
\end{figure}

If these neutrino sources are optically thin, then the upper bound on
neutrino flux is obtained from the well-measured cosmic-ray flux,
because each proton that arrives at Earth should produce no more than a
few neutrinos at the source.
Based on this argument and assuming that the cosmic-ray spectrum above
$10^{9}$~GeV is of extra-galactic origin, Waxman and
Bahcall~\cite{WBBound} derived the upper bound for the $\nu_\mu$ flux
before neutrino oscillation to be $E_{\nu_\mu}^2 \rmd \Phi_{\nu_\mu}^{\rm
WB}/ \rmd E_{\nu_\mu} = (1$--$4) \times 10^{-8}$ GeV cm$^{-2}$ s$^{-1}$
sr$^{-1}$, where the range reflects cosmological evolution of source
density.
As we expect that flavor ratio at production ({\em i.e.}, before
oscillation) is $\nu_e$:$\nu_\mu$:$\nu_\tau = 1$:2:0 due to meson
decays, the WB bound summed over flavors after neutrino oscillation is
$E_\nu^2 \rmd\Phi_\nu^{\rm WB}/\rmd E_\nu = (1.5$--$6) \times 10^{-8}$ GeV
cm$^{-2}$ s$^{-1}$ sr$^{-1}$.
Here we adopt $E_\nu^2 \rmd\Phi_\nu^{\rm WB}/\rmd E_\nu = 5 \times 10^{-8}$ GeV
cm$^{-2}$ s$^{-1}$ sr$^{-1}$ for our reference value and show this bound
in \Fref{flux}.

We stress that although all previous studies adopted the WB upper
limit on astrophysical neutrino flux as their incident source, that
flux is not a model {\em prediction}, but an {\em upper bound}.  One
can never regard the output of a WB maximal neutrino flux as a solid
prediction for the stau event rate at IceCube; the resulting stau flux
is merely an upper limit.
In addition, a few comments are in order on the WB bound's robustness,
especially at our energies of interest.
First, the WB bound is valid for $E_\nu > 5\times 10^{7}$~GeV, the
threshold corresponding to proton energies of $10^{9}$~GeV (a daughter
neutrino carries $\sim 5\%$ of its parent proton's energy~\cite{WBBound}).
Below this, the WB bound is only an {\it extrapolation}, since
the cosmic-ray flux below $10^{9}$~GeV is totally dominated by the
galactic component.
Second, by its definition the WB bound is applicable only to
optically thin sources.
If neutrino-emitting opaque objects existed, their
contributions might sum to produce a neutrino flux exceeding the WB 
bound.  Potential sources include baryon-rich GRBs~\cite{ABJet} and 
starburst galaxies~\cite{starburst}.

We also adopt a model for high-energy neutrino production via shocks
in GRBs~\cite{WBGRB}.  In contrast to the WB bound, it is a
prediction; see the red dashed line in \Fref{flux}.

\subsection{Atmospheric conventional neutrinos}

While the astrophysical neutrino flux is totally unknown, there is a
guaranteed and well-measured neutrino component---atmospheric
neutrinos.  These arise from the decays of mesons produced by cosmic
rays striking the upper atmosphere.  Neutrinos coming from pion and
kaon decays form the `conventional' component, which is well-studied
both theoretically~\cite{AtmTheory} and experimentally~\cite{AtmExp}.
Although there is no detection of any neutrinos for $E_\nu \gtrsim
10^5$ GeV, the energy range we are mainly interested in, this
component should quite easily be extrapolated using measured data at
lower energies, thus providing guaranteed seeds for CHAMP production.
We use the model of Ref.~\protect\cite{Candia:2003ay} for the
conventional atmospheric flux.  As shown in \Fref{flux}, the
well-known spectrum of these neutrinos falls rather steeply with
increasing energy.

\subsection{Atmospheric prompt-decay neutrinos}

Hadronic cosmic ray interactions with Earth's atmosphere also produce
short-lived charmed
mesons~\cite{Zas:1992ci,Pasquali:1998ji,Gelmini:1999xq,Costa:2000jw,Volkova:2001th,Fiorentini:2001wa,Beacom:2004jb,Martin:2003us,Achterberg:2006pw}.
Even though branching ratios into these final states are not large,
the neutrino spectrum from the subsequent decays of charmed mesons is
quite hard as they immediately decay, before losing energy.  As a
consequence, the contribution from this `prompt-decay' component to
the total flux of the atmospheric neutrinos falls less rapidly with
energy than the conventional component.  The absolute normalization
of this prompt flux is however still unknown, and there is a large
range in model predictions.  The main source of uncertainty is the
proper treatment of next-to-leading order charm meson production cross
sections, which strongly depend on the behavior of the nucleon parton
distribution functions (PDFs).  In our study, we consider a range
between a smaller prompt flux from Ref.~\cite{Beacom:2004jb} and a
larger prompt flux obtained by using the shape presented in
Ref.~\cite{Martin:2003us} and normalizing it to IceCube's experimental
upper limit~\protect\cite{Achterberg:2006pw}.  Such a large flux, just
allowed by the data, is in fact characteristic of the largest model
predictions among
Refs.~\cite{Zas:1992ci,Pasquali:1998ji,Gelmini:1999xq,Costa:2000jw,Volkova:2001th,Fiorentini:2001wa,Beacom:2004jb,Martin:2003us,Achterberg:2006pw}.
We show these fluxes in \Fref{flux}.


\section{Interaction cross section}\label{sec:particle}

Several TeV-scale extensions of the Standard Model feature a
meta-stable massive charged particle.  Perhaps the best-motivated
scenario from a theoretical standpoint is supersymmetry, which
provides several examples.  If the LSP is very weakly
interacting---{\em e.g.}, a gravitino or a right handed sneutrino,
where the interactions with the rest of the supersymmetric partners
are suppressed by gravitational couplings or a gauge symmetry---the
NLSP is generally meta-stable.  Specifically, a charged NLSP can occur
in supergravity theories with a gravitino
LSP~\cite{Blumenthal:1982mv,Feng:2004mt,sugra}, gauge mediated
supersymmetry breaking setups~\cite{gmsb,Giudice:1998bp}, scenarios
featuring a stau--neutralino near-degeneracy (particularly in the
so-called co-annihilation region~\cite{coan,Profumo:2004qt}; here one
can have a neutralino LSP), or supergravity scenarios with a
right-handed sneutrino LSP~\cite{rhnlsp}.

Another TeV-scale new physics setup that naturally encompasses a
meta-stable NLP is that of universal extra-dimensions (UED) with a
Kaluza-Klein (KK) graviton as the lowest mass eigenstate in the KK
tower~\cite{Hooper:2007qk}. In the minimal UED setup, the
next-to-lightest KK state is usually neutral, and corresponds to the
KK first excitation of the U(1) gauge boson, $B^{(1)}$, if the Higgs
mass is below $\approx$200 GeV, for any value of the compactification
inverse radius $R^{-1}$. However, as pointed out
in Ref.~\cite{Cembranos:2006gt}, even in the minimal UED setup the
next-to-lightest KK particle (NLKP) can be charged, and the LKP can be
the KK graviton if $R^{-1}\lesssim 809$ GeV {\em and} $m_h\gtrsim 250$
GeV. In this case, the NLKP corresponds to the KK first charged Higgs
mode. Alternatively, the boundary conditions at the orbifold fixed
points can alter the spectrum, and give rise to scenarios with a KK
lepton as the NLKP, and again, a KK graviton LKP
\cite{Hooper:2007qk}. In general, electroweak precision observables
constrain the scale $R^{-1}$ where the first excitation of a 
5-dimensional UED scenario might be expected to a few hundred GeV,
depending upon the value of the Higgs mass~\cite{uedewpo}. The
analysis outlined below applies, with the appropriate production cross
sections and energy losses, to such UED models and to any other
similar framework featuring a meta-stable charged particle.

We choose to work with two of the well-known and well-motivated
supersymmetric frameworks mentioned above: gauge mediated
supersymmetry breaking (GMSB)~\cite{gmsb,Giudice:1998bp}, and minimal
supergravity (mSUGRA) with a gravitino LSP (see, {\em e.g.},
Refs.~\cite{Blumenthal:1982mv,Feng:2004mt,sugra}).  For each framework
we examine two models.  One is a `supersymmetric benchmark' (the SPS7
point of Ref.~\cite{Allanach:2002nj} for GMSB, and the $\varepsilon$
model of Ref.~\cite{DeRoeck:2005bw} for mSUGRA with gravitino LSP),
and the other is a variant with a lighter spectrum (models {\bf I} and
{\bf II}).  These are essentially rescalings of the first two.  By
adopting benchmark models, not only do we consider phenomenologically
viable and theoretically soundly-motivated setups, but we also make it
easier to compare the detection technique discussed here with a wealth
of existing phenomenological analyses of the same models (see, {\em
e.g.}, Refs.~\cite{Allanach:2002nj,DeRoeck:2005bw}).  As will become
apparent below, for the present analysis the details of the spectrum
of the heavy supersymmetric particle pair produced is crucial.
Assuming a degenerate sfermion spectrum, while potentially useful to
get an understanding of the role of the mass scale in the expected
size of the signal, can potentially be a misleading
over-simplification.

\begin{table}[!ht]
\begin{center}
\vspace*{0.4cm}
\begin{tabular}{|c|c|c|c|c|c|}\hline
mSUGRA Models 
        & $M_{1/2}$ & $m_0$ & $\tan\beta$  & ${\rm sgn}(\mu)$ & $A_0$\\
\hline
{\bf I} & 280 GeV & 10 GeV & 11 & $>0$ & 0\\
{\boldmath{$\varepsilon$}} \cite{DeRoeck:2005bw}
        & 440 GeV & 20 GeV &
 15 & $>0$ &
 $-25$ GeV\\
\hline
\hline
GMSB Models 
         & $M_{\rm mes}$ & $\Lambda$ & $\tan\beta$  & ${\rm sgn}(\mu)$ & $N_{\rm mes}$\\
\hline
{\bf II} & 70 TeV & 35 TeV & 15 & $>0$ & 3\\
{\bf SPS7} \cite{Allanach:2002nj} 
         & 80 TeV & 40 TeV & 15 & $>0$ & 3\\
\hline
\end{tabular}
\end{center}
\caption{The input parameters for the mSUGRA (upper pair) and GMSB
(lower pair) models used in the present analysis and in
\Fref{xsec}.}
\label{tab:models}
\end{table}
\begin{table}[!ht]
\begin{tabular}{|c|c|c|c|c|}\hline
Model & $m_{\wt\tau_1}$ & $m_{\wt q_1}$ & $m_{\wt \chi^0_1}$  & $m_{\wt \chi^\pm_1}$ \\
\hline
{\bf I} & 101 GeV & 620 GeV & 110 GeV & 200 GeV\\
\boldmath{$\varepsilon$} \cite{DeRoeck:2005bw}
      & 153 GeV & 940 GeV & 180 GeV & 340 GeV\\
\hline
\hline
{\bf II} & 101 GeV & 800 GeV & 140 GeV & 240 GeV\\
{\bf SPS7} \cite{Allanach:2002nj} & 120 GeV & 900 GeV & 160 GeV & 270 GeV\\
\hline
\end{tabular}\centering
\caption{The masses of the lightest stau, first-generation squark,
and lightest neutralino and chargino, for the four models of 
\Tref{tab:models}.}
\label{tab:spectrum}
\vspace*{0.4cm}
\end{table}

We specify the mSUGRA and GMSB input parameters in
\Tref{tab:models}. Notice in particular that Model {\bf II} lies on
the SPS7 slope defined in Ref.~\cite{Allanach:2002nj}.  In
\Tref{tab:spectrum} we detail the four models' relevant particles
masses.

The gravitino mass need not be specified as long as the stau decay
length, $c\tau_{\wt\tau}$, is larger than or of the same order as the
Earth radius, $R_\oplus$.  This implies a lower limit on the
gravitino mass $m_{\wt G}$~\cite{Ellis:2003dn},
\begin{equation}
c\tau_{\wt\tau} \simeq 1.7 \times 10^9\ {\rm km}
\left(\frac{m_{\wt G}}{100\ {\rm GeV}}\right)^2
\left(\frac{1\ {\rm TeV}}{m_{\wt\tau}}\right)^5
\left(1-\frac{m_{\wt G}^2}{m_{\wt\tau}^2}\right)^{-4}
\gtrsim R_\oplus
\: .\label{eq:lifetime}
\end{equation}
Rearranged, the formula implies, for instance, that for a 100~GeV stau
the gravitino mass can be as light as 1~MeV, and for a 1~TeV stau
$m_{\wt G}$ has to be larger than 0.3~GeV. Notice that
\Eref{eq:lifetime} does not take into account the relativistic boost
factor, $\gamma$, which is sizable for staus produced in very high
energy neutrino--nucleon interactions.  It is therefore a conservative
constraint on the gravitino mass.

In addition, the stau lifetime should be short enough to be consistent
with limits obtained from the effects on light element abundances processed in
big-bang nucleosynthesis~\cite{BBN} (see also Refs.~\cite{BBN2}) and
from excessive distortions to
the cosmic microwave background spectrum~\cite{CMB}.  Of particular
relevance are constraints resulting from overproduction of ${}^6$Li
and ${}^7$Li~\cite{BBNLDP,pospelov,Kaplinghat:2006qr}, induced by catalytic
effects produced by bound states consisting of light nuclei and the
meta-stable CHAMP (here, the lightest stau). Notice, however, that the
details of the estimate of the amount of primordially synthesized
lithium are still under debate~\cite{jedamzik}.  Also, the constraints
depend upon the fraction of electromagnetic energy released in the
decay. Conservatively, if one requires the lifetime of the charged
meta-stable species to be shorter than $10^3$--$10^4$ s, as implied by
the analysis of Ref.~\cite{olive}, the gravitino mass is constrained
to be approximately below 1 GeV for a 100 GeV stau, and below 100 GeV
for a 1 TeV stau.  This evidently leaves a very wide window, of almost
three orders of magnitude, for the viable gravitino mass range.

To calculate the stau flux,\footnote{Here the word `stau' denotes staus
and anti-staus collectively.  This is because both could be produced by
the same interaction, but cannot be distinguished at neutrino
telescopes.}
we first need to calculate the stau
production cross section as a function of incoming neutrino energy.
We do this using the {\sc susy-madevent}
package~\cite{Cho:2006sx,Maltoni:2002qb}, which calculates the
differential or total cross section for any $2\to n$ scattering
process in the MSSM given an SLHA-conforming (standardized format for
spectra) model input file~\cite{SLHA}.  We generate model input files
using the {\sc suspect} spectrum generator
package~\cite{Djouadi:2002ze}, which also automatically checks the
generated model against various known precision data constraints, such
as $b\to s\gamma$.  Our four models do not conflict with any known
constraints.
The processes contributing to stau production mainly stem from
tree-level $u$ and $t$ exchange diagrams with a slepton ($\tilde l$ and
$\tilde \nu$) plus a squark ($\tilde u$ and $\tilde d$) in the final
state, through neutralino ($\chi^0$) or chargino ($\chi^\pm$) exchange,
as discussed in Ref.~\cite{Ahlers:2006pf}.
More specifically, they are: $\nu + u (d) \stackrel{\chi^\pm}{\to}
\tilde l + \tilde d (\tilde u)$, $\nu + u (d) \stackrel{\chi^0}{\to}
\tilde \nu + \tilde u (\tilde d)$, where $\chi^0$ and $\chi^\pm$ indicate
neutralinos and charginos, respectively.
The squarks and sleptons produced in the final state promptly cascade
decay into meta-stable staus.

We calculate total cross sections, summing over neutrino and
anti-neutrino inelastic scattering on protons using exact matrix
elements to supersymmetric pair final states for both charged current
(CC) and neutral current (NC), employing CTEQ6L1
PDFs~\cite{Pumplin:2002vw}.  The possible final states are
(anti-)sneutrino or (anti-)slepton, plus (anti-)squark.  Our
calculations are leading order, as there are no available NLO QCD
corrections for neutrino--proton ($\nu p$) scattering.  Judging from
the known results for proton--proton ($pp$) scattering, however, we
probably make an underestimate of the rate on the order of $50\%$.  In
this regard our calculation is conservative.  For our purposes at the
relevant energies, the neutrino--neutron cross section is sufficiently
close to the neutrino--proton cross section that we can ignore 
calculating it separately.

An interesting side observation is that the $\nu p$ and $\bar\nu p$
cross sections are not equal except at very large $\sqrt{s}$, where
low-$x$ quarks dominate the PDFs and are approximately
egalitarian~\cite{Carena:1998gd}; we reproduce this observation here.
As $\sqrt{s}$ gets within a couple orders of magnitude above threshold,
however, $\bar\nu p$ dominates because the larger CC process picks out
valence quarks, where $u$ dominates slightly over $d$.  Closer to
threshold this remains true for the CC process, but for NC $\nu p$
dominates because of chirality-selection for the final-state squarks.
Near threshold these two diverging components accidentally roughly
cancel, resulting in $\sigma_{\nu p}\approx\sigma_{\bar\nu p}$ once
again.  For the present purposes, however, we are concerned only with
total rates, and ignore charge-separated subsamples, which would vary
somewhat as a function of stau energy.

\begin{figure}[t]
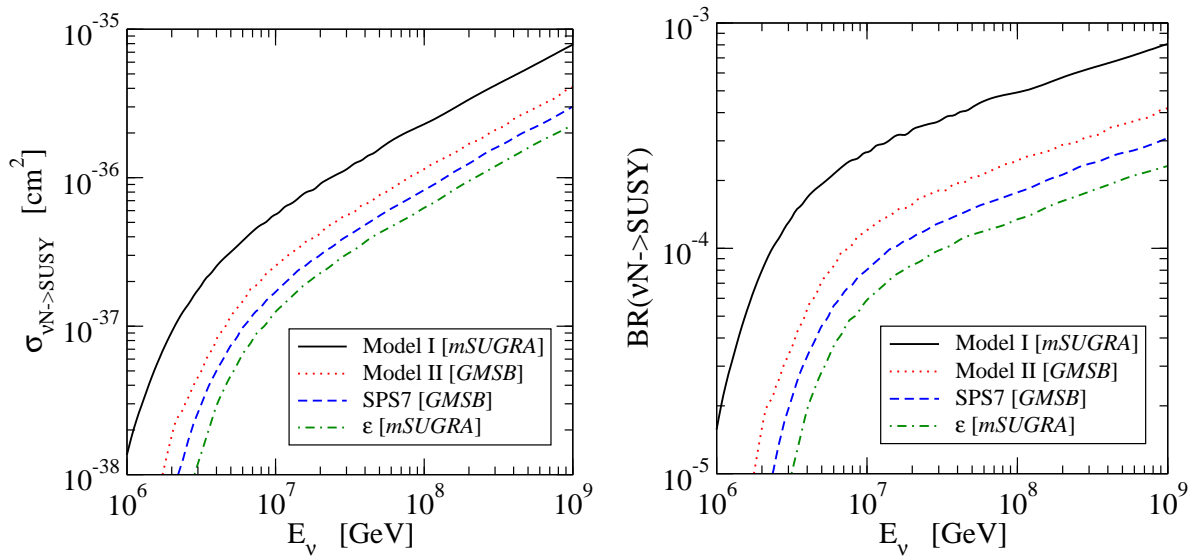

\begin{center}
\mbox{\includegraphics[width=7.8cm,clip]{figures/xsec_def}\quad
\includegraphics[width=7.5cm,clip]{figures/ratios}}
\caption{{\em Left:} The neutrino--proton cross sections, as a
  function of the incident neutrino energy for the production of stau
  pairs in the four models under consideration here (see text and
  tables~\ref{tab:models} and \ref{tab:spectrum} for details on the
  specific models). {\em Right:} The ratio ${\rm BR}(\nu{\rm N}\to{\rm
  SUSY})$ of the neutrino--proton cross section into supersymmetric
  particle pairs over the total neutrino--proton cross section, as a
  function of the incident neutrino energy, for the same four models
  as in the left panel.}
\label{xsec}
\end{center}
\end{figure}

The left panel of \Fref{xsec} illustrates our results for the
neutrino--proton scattering cross section $\sigma_{\nu p}(E_\nu)$ into
any supersymmetric particle pairs, for the four models listed in
\Tref{tab:models}, as a function of the incident neutrino energy. The
general trends in $\sigma_{\nu p}(E_\nu)$ are consistent with those
found in other analyses, see {\em e.g.}, Ref.~\cite{Ahlers:2006pf}: the
steep rise in the low-energy end reflects the strong kinematic
suppression associated with the squark--slepton pair production
threshold. The subsequent rise of the cross section with the incoming
neutrino energy depends upon the small-$x$ behavior of the PDFs. We
give in \Sref{sec:susy}, where we discuss the role of the specific
supersymmetric particle spectrum in the determination of $\sigma_{\nu
  p}(E_\nu)\to{\rm SUSY}$, an analytical interpretation of the
specific power-law behavior that emerges from the numerical
computation.
Notice that compared to the optimistic toy model used in
Ref.~\cite{Ahlers:2006pf}, theoretically-motivated (optimistic)
supersymmetric setups appear to give a maximal $\sigma_{\nu p}$ that
is roughly one order of magnitude smaller in the asymptotic
high-energy regime.

In the right panel of \Fref{xsec}, we show the ratio of the
neutrino--nucleon cross section into supersymmetric particle pairs over
the total neutrino--nucleon cross section (that is, as apparent from
the figure, always close to the purely Standard Model cross
section). As we explain in \Sref{sec:thicktarget}, this is the
physical quantity of interest, in the limit of Earth as a thick
target for high energy neutrinos, for the computation of the stau
flux. Beyond threshold effects, we point out that in the energy range
of interest the branching ratio of neutrino--nucleon interactions into
supersymmetric particle pairs lies between $10^{-4}$ and $10^{-3}$.

\section{Flux of long-lived staus}\label{sec:rates}

We devote this section to a detailed analytical treatment of the
computation of the flux of staus produced by neutrino--nucleon
collisions that might be detected at a neutrino telescope such as
IceCube.  We start, in \Sref{sub:formulation}, with a derivation of
the differential flux of staus from first principles, leading to the
result presented in \Eref{eq:detection flux}.  In the following
\Sref{sec:thicktarget} we assume that Earth is opaque to neutrinos at
energies relevant here.
In this {\em thick
target approximation}, we analytically show that the flux of staus at
the detector can be computed as a simple integral, over incident
neutrino energies, shown in \Eref{eq:stau flux: simple}, of the product of three factors:
\begin{enumerate}
\item the differential flux of incident neutrinos
(shown in \Fref{flux}),
\item the ratio of the neutrino--nucleon cross
section into supersymmetric particles over the total neutrino--nucleon
cross section (shown in \Fref{xsec}, right), and 
\item a `geometric
efficiency' factor, to be defined below and explicitly shown in the
right panel of \Fref{fig:geoeff}.
\end{enumerate}

When a neutrino interaction occurs, the branching ratio for stau
production among the final states is given by the above cross section
ratio.  In the thick-target approximation, all incoming neutrinos below
the horizon will interact in Earth.  Since the staus are collinear with
the incoming neutrino direction, and are produced with a sizable
fraction of the neutrino energy, this leads to a simple but important
result about the stau flux.  Neglecting stau energy losses in matter
for a moment, we see that Earth acts as a neutrino-to-stau converter,
with a probability that is independent of direction (below the horizon).
This gives an upper bound on the stau flux through IceCube.
The relevant range of energies will be between threshold ($\simeq
10^6$ GeV; see \Fref{xsec}) and the point at which the stau flux
becomes too small ($\simeq 10^7$ GeV; note the product of
Figures \ref{flux} and \ref{xsec}).  Taking stau energy losses into account will
only reduce the stau flux at the detector.  Since staus cannot reach
the detector from too far away, this means that only a limited range
of nadir angles will be relevant, and this defines our geometric
efficiency.

With our approximations,
we give a preliminary assessment of the total expected
stau flux at the detector (\Tref{tab:approx}). Finally, we compute the
actual accurate stau flux resulting from the full-glory integration of
\Eref{eq:detection flux} in \Sref{sub:integra}. We provide
numerical estimates of the integrated flux in \Tref{tab:true} (that
fall within a factor 2 of the approximate results anticipated in
\Tref{tab:approx}), as well as the actual differential flux of staus
from the different primary incident neutrino sources. Finally, we
comment in \Sref{sec:NN} on the flux of staus predicted from
nucleon--nucleon reactions.

\subsection{Formulation}\label{sub:formulation}

In the framework we consider here, all supersymmetric particles
produced in neutrino--nucleon interactions promptly decay into the
NLSPs---here, the stau, which is meta-stable: long-lived enough to
propagate through Earth (with energy loss) and reach the detector.
Our objective is to calculate the spectrum of staus after this energy
loss, following similar principles for the spectra of neutrino-induced
muons~\cite{Gaisser:1985cm,Kistler:2006hp}.

The stau electromagnetic energy loss rate is given by
\cite{Albuquerque:2003mi,Reno:2005si}
\begin{equation}
 \frac{\rmd E_s}{\rmd X} = \alpha_s + \beta_s E_s,
  \label{eq:dEdX}
\end{equation}
where $X = \rho\times l$ is the column depth of matter in units of
g$\cdot$cm$^{-2}$ ({\em i.e.}, density times distance), and the
$\alpha_s$ and $\beta_s$ terms represent ionization and radiation
losses, respectively.  We neglect discrete scattering by weak
interactions, as the effect is small at the energies we focus on near
threshold~\cite{Huang:2006ie}.  Hereafter, we use the subscript $s$ to
indicate quantities referring to staus.  Our coordinate system locates
the detector at $X = 0$ and particles are produced at $X > 0$, so that
the energy $E_s$ is a growing function of $X$ ($\rmd E_s/\rmd X$ is
positive).  For the density profile of Earth, we use the model
given in Ref.~\cite{Gandhi:1995tf}.

The ionization coefficient for staus, $\alpha_s$, is approximately the
same as that of muons~\cite{Reno:2005si}; specifically, we use
$\alpha_s = 2 \times 10^{-3}$ GeV cm$^2$ g$^{-1}$.  Radiative losses,
on the other hand, depend on particle mass, and we take the corresponding
coefficient to be given by $\beta_s = 4.2
\times 10^{-6} (m_\mu/m_s)$ cm$^2$ g$^{-1}$~\cite{Reno:2005si}.  By
integrating \Eref{eq:dEdX}, the `distance' $X_{if} = X_i - X_f$
traversed by the stau while its energy decreases from $E_s^i$ to
$E_s^f$ reads
\begin{equation}
X_{if} = \frac{1}{\beta_s} \ln\left(\frac{\alpha_s + \beta_s E_s^i}
{\alpha_s + \beta_s E_s^f}\right).
\label{eq:X_if}
\end{equation}
We assume that each stau produced in $\nu p$ interactions carries a
large fraction of the parent neutrino energy.  We denote this as
$E_s^i = (1-y) E_\nu$ and assume $y = 0.5$, independent of neutrino
energy.
(We discuss in \Sref{sec:thicktarget} below the dependence of the stau flux on the parameter $y$.)
At these high laboratory energies, the staus may be taken to be collinear
with the original neutrino direction.
The differential flux of produced staus per energy $E_s^i$
and distance $X$ reads
\begin{equation}
\fl
 h_s(E_s^i) = \frac{\rmd^2 \Phi_s^i}{\rmd E_s^i \rmd X} = 
  \frac{\sigma_{\rm SUSY}(E_\nu)}{(1-y)m_p}
 \frac{\rmd\Phi_\nu(E_\nu)}{\rmd E_\nu}
  \exp\left[-\frac{(X_{\rm max}-X)\sigma_{\rm tot}(E_\nu)}{m_p}\right],
\end{equation}
where $E_\nu = E_s^i / (1-y)$.
The exponential factor takes into account the neutrino attenuation,
mainly due to the Standard Model interactions ($\sigma_{\rm tot}
\approx \sigma_{\rm SM}$; see \Fref{xsec}, right); $X_{\rm max}$
represents the value of $X$ corresponding to the surface of Earth,
which depends on the direction.
The overall $(1-y)^{-1}$ factor comes from the change of variables
from the spectrum of neutrinos to that of produced staus, {\em i.e.},
$\rmd E_\nu / \rmd E_s^i = (1-y)^{-1}$.

The spectrum of staus at the detector is given by a double integral
over all production positions and energies, subject to the constraint of
having the stau energy be between $E_s^f$ and $E_s^f + dE_s^f$ at $X =
0$:
\begin{equation}
 \frac{\rmd\Phi_s^f}{\rmd E_s^f} = \int_{E_s^f}^\infty \rmd E_s^i
  \int_0^\infty \rmd X\, h(E_s^i)
  \delta\left(E_s^f - f(E_s^i, X)\right),
\end{equation}
where the function $f(E_s^i, X)$ is defined by the energy loss,
\Eref{eq:X_if}.  We use the energy constraint to perform the $E_s^i$
integration, and as a result we obtain the following expression:
\begin{eqnarray}
 E_s^f\frac{\rmd\Phi_s^f}{\rmd E_s^f} & = &
 \int_0^{X_{\rm max}} \rmd X
 \left[\frac{\exp[-(X_{\rm max} - X) \sigma_{\rm SM}(E_\nu) / m_p]}
  {m_p / \sigma_{\rm SM}(E_\nu)}\right] \nonumber \\
 &  & \times
  \left[\left(\frac{E_s^f}{E_s^i}\right) \left(E_\nu
	 \frac{\rmd\Phi_\nu}{\rmd E_\nu}\right)
  \left(\frac{\sigma_{\rm SUSY}(E_\nu)}{\sigma_{\rm SM}(E_\nu)}\right)
  e^{\beta_s X} \right],
  \label{eq:detection flux}
\end{eqnarray}
where $E_\nu$ and $E_s^i$ inside the integral have to be evaluated
according to the chosen $E_s^f$ on the left-hand side and the $X$ at
that step inside the integral.  From the energy-loss equation,
\Eref{eq:X_if}, we have
\begin{equation}
E_s^i = \left(E_s^f + \frac{\alpha_s}{\beta_s}\right)
  e^{\beta_s X} - \frac{\alpha_s}{\beta_s},
\label{eq:energy relation}
\end{equation}
where, again from the kinematic definition, $E_\nu = E_s^i / (1 - y)$.

It is convenient to change the integration variable by dividing the
differential $\rmd X$ by $X$ and multiplying the integrand by $X$.
Then the integration steps are in $\ln X$, and in \Fref{fig:xdist} we show
this new integrand for different nadir angles.
In the left-hand panel, the stau energy losses in matter are neglected,
so that the neutrino interaction and geometric effects are shown clearly.
For each nadir angle ($90^\circ$ is at the horizon, and $0^\circ$ is along
the diameter of Earth), the sharp edges at large $X$ occur due to the
boundary of Earth.  The peaks arise because the neutrino interactions
occur logarithmically near $X = X_{\rm max}$.  Substantial attenuation of
the neutrino flux is only seen at small nadir angles; in units of the 
axis, the exponential scale height is $\simeq 0.1$.  In the right-hand
panel, the stau energy losses are now included.  As expected, this
prevents staus from arriving at the detector from too large of distances
(beyond $\simeq 0.01$--0.05 in units of the axis, depending on 
nadir angle due to the radial variation of the density profile).
In this panel, the visual area under each curve shows its relative
importance to the total stau flux through the detector.

\subsection{Thick target approximation}
\label{sec:thicktarget}

For neutrino energies relevant for our purposes, $E_\nu \gtrsim 10^6$
GeV, Earth is opaque to neutrinos at the most important nadir angles.
Thus, to a good approximation,
we can assume that staus are produced from neutrino interactions
logarithmically near Earth's surface.
In this case, the exponential factor in
equation~(\ref{eq:detection flux}) is sharply peaked at $X = X_{\rm max}$,
{\em i.e.}, $\exp[-(X_{\rm max}-X)\sigma_{\rm SM}/m_p] / (m_p /
\sigma_{\rm SM}) \simeq \delta (X - X_{\rm max})$.
For our change of variables, the integrand is proportional to $\delta
(\ln X - \ln X_{\rm max}) = X \delta (X - X_{\rm max})$.

We then obtain
\begin{equation}
\fl
E_s^f \frac{\rmd\Phi_s^f}{\rmd E_s^f} \simeq
\frac{E_s^f}{E_s^i} \left(E_\nu \frac{\rmd\Phi_\nu}{\rmd E_\nu}\right)
\frac{\sigma_{\rm SUSY}}{\sigma_{\rm SM}} e^{\beta X_{\rm max}}
 =
\frac{E_s^f}{E_s^i} \frac{\rmd E_s^i}{\rmd E_s^f}
\left(E_\nu \frac{\rmd\Phi_\nu}{\rmd E_\nu}\right)
\frac{\sigma_{\rm SUSY}}{\sigma_{\rm SM}},
\end{equation}
where, in the second equality, we used the relation $\rmd E_s^i/\rmd
E_s^f = e^{\beta X_{\rm max}}$ from \Eref{eq:energy relation} with $X
= X_{\rm max}$.

\begin{figure}
\begin{center}
\includegraphics[width=7.5cm]{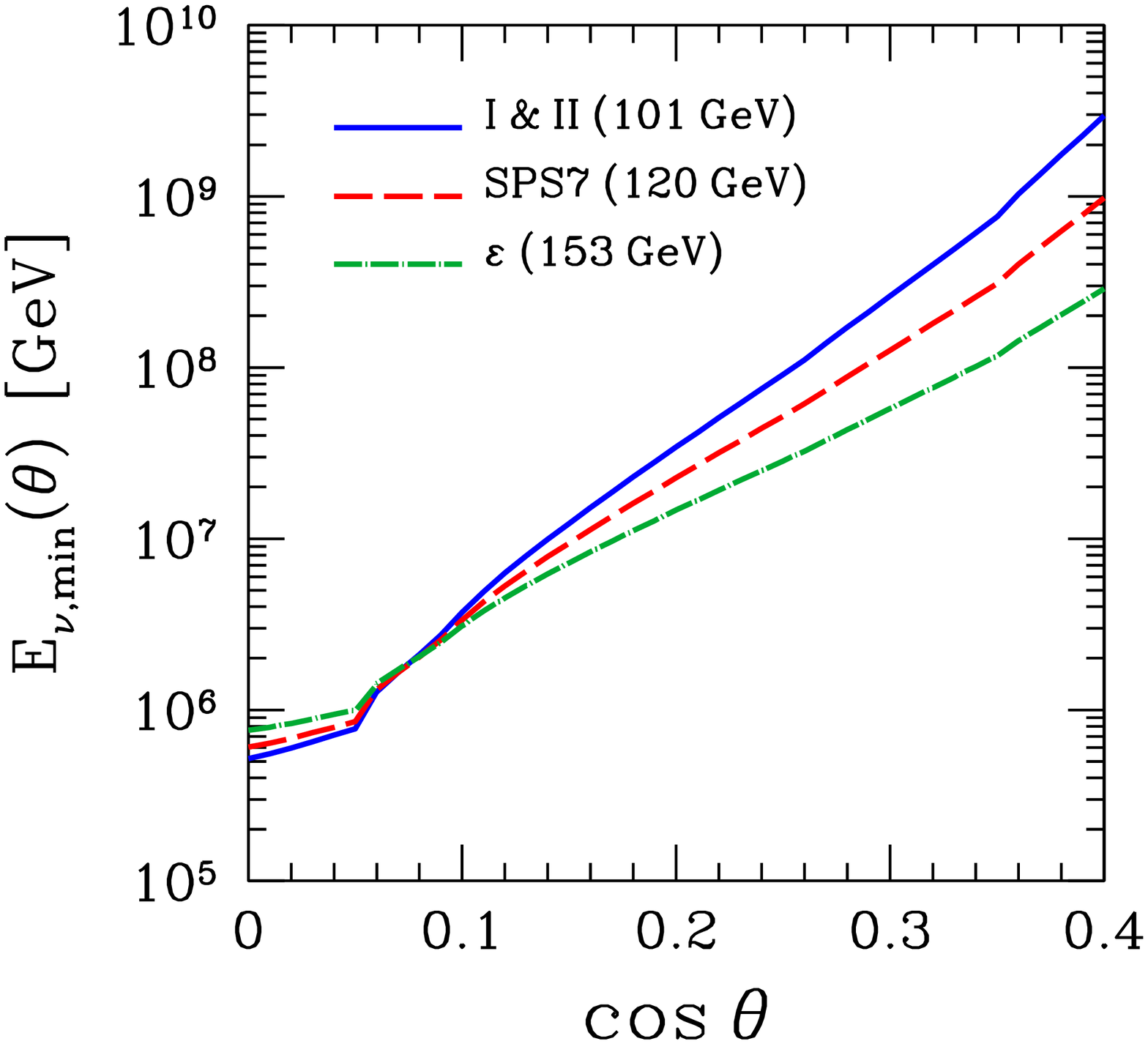}
\includegraphics[width=7.5cm]{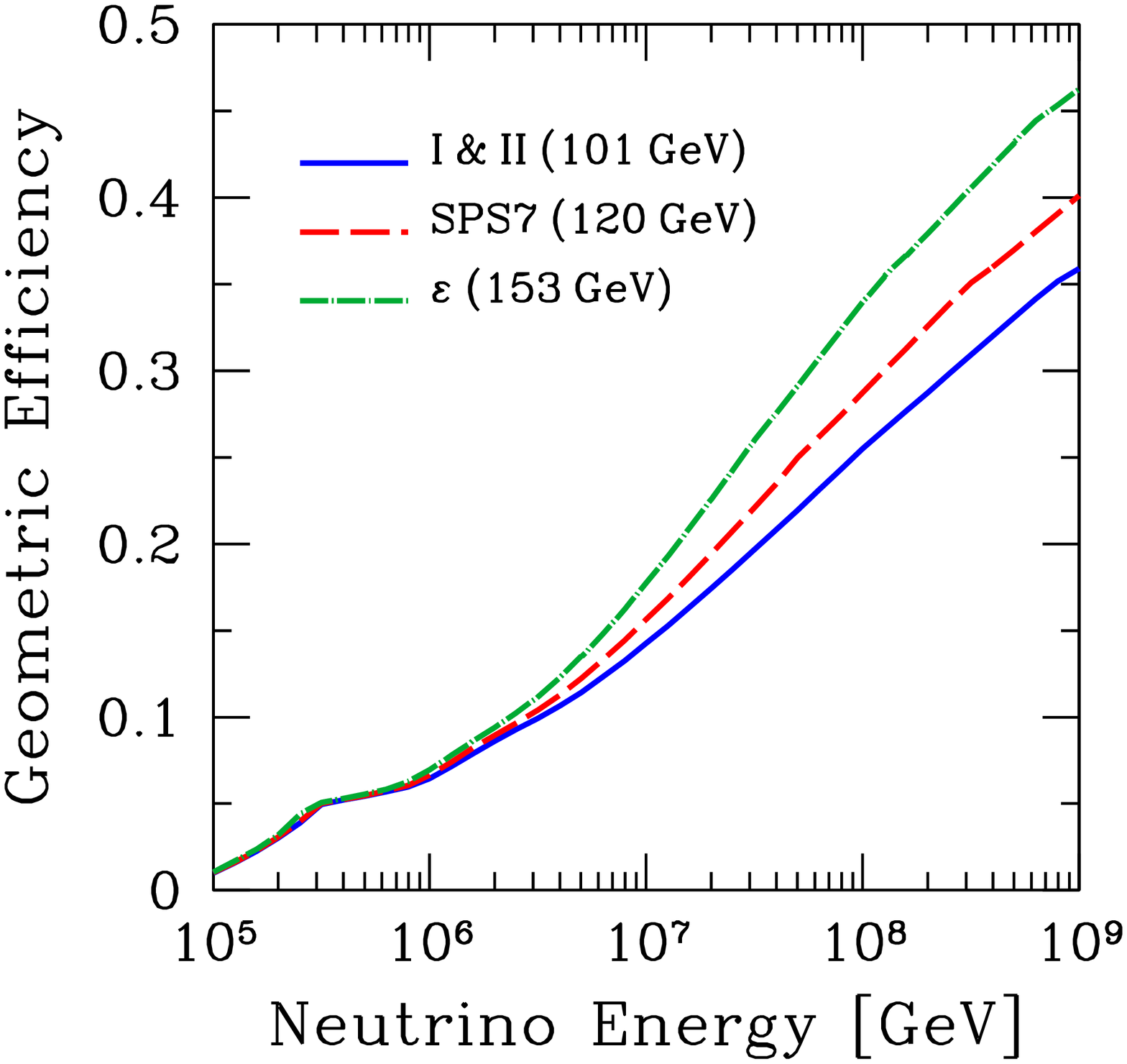}
\caption{{\em Left:} Minimum neutrino energy, $E_{\nu,{\rm
  min}}(\theta)$, below which the produced staus cannot reach the
  detector, as a function of the nadir angle $\theta$.  {\em Right:}
  The geometric efficiency factor $\epsilon_{\rm geo}(E_\nu)$, as a
  function of the incident neutrino energy (see \Eref{eq:stau
  flux: simple}).}
\label{fig:geoeff}
\end{center}
\end{figure}

We require the staus to be more energetic than a given detector
threshold $E_{s,{\rm th}}$.  If the staus are relativistic enough to emit
\v Cerenkov light, they will be detected.  We adopt $E_{s,{\rm th}} =
300$ GeV, since the typical stau mass we consider is $\sim$100 GeV,
but we note that our results change negligibly for higher or lower
thresholds.  For instance, a shift by one order of magnitude,
$E_{s,{\rm th}} = 3$ TeV, affects the final result by only $0.2\%$.

To satisfy the detection requirement $E_s^f > E_{s,{\rm th}}$, the
initial stau and corresponding neutrino energies must be larger than
some minima, $E_{s,{\rm min}}^i$ and $E_{\nu,{\rm min}}$, given by
\begin{equation}
\fl
E_{s,{\rm min}}^i(\theta) =
\left(E_{s,{\rm th}} + \frac{\alpha_s}{\beta_s}\right)
e^{\beta_s  X_{\rm max}(\theta)} - \frac{\alpha_s}{\beta_s},
~~ E_{\nu,{\rm min}}(\theta) = \frac{E_{s,{\rm min}}^i(\theta)}{1-y},
\end{equation}
as evident from \Eref{eq:energy relation}.  We plot this
minimal neutrino energy as a function of the nadir angle $\theta$ in
the left panel of \Fref{fig:geoeff} for the various supersymmetry
models under consideration.  Note that the crucial quantity here is
the stau mass, hence we obtain the same result for models {\bf I} and
{\bf II}.  In the thick target approximation, the flux of staus
reaching the detector from below is thus given by
\begin{eqnarray}
F_s & = & \int_{\theta < \pi/2} \rmd\Omega
\int_{E_{s,{\rm th}}}^\infty \rmd E_s^f \frac{\rmd\Phi_s^f}{\rmd
  E_s^f}
\nonumber\\ &=&
\int_{\theta < \pi /2} \rmd\Omega
\int_0^\infty \rmd E_\nu \left(\frac{\rmd\Phi_\nu}{\rmd E_\nu}\right)
\left(\frac{\sigma_{\rm SUSY}}{\sigma_{\rm SM}}\right)
\Theta\left(E_\nu - E_{\nu,{\rm min}}(\theta)\right)
\nonumber\\ &\equiv&
2\pi \int_0^\infty \rmd E_\nu \left(\frac{\rmd\Phi_\nu}{\rmd
  E_\nu}\right)
\left(\frac{\sigma_{\rm SUSY}}{\sigma_{\rm SM}}\right) \epsilon_{\rm
geo}(E_\nu),
\label{eq:stau flux: simple}
\end{eqnarray}
where $\Theta$ is the step function, and in the second equality, we have
simply used definitions given above.  In the last equality, we are
defining a `{\em geometric efficiency}' factor $\epsilon_{\rm
  geo}(E_\nu)$, assuming that the incident neutrino intensity
$\Phi_\nu$ is isotropic. We show this geometric efficiency factor in
the right panel of \Fref{fig:geoeff}.
Thus, under the thick target approximation, the detection flux of
staus can be divided into three independent factors: 
\begin{itemize}
\item[$\bullet$] incident neutrino spectrum $\rmd\Phi_\nu/\rmd E_\nu$
\item[$\bullet$] cross section ratio $\sigma_{\rm SUSY}/\sigma_{\rm SM}$
\item[$\bullet$] geometric efficiency $\epsilon_{\rm geo}$
\end{itemize}
These three factors are illustrated in
figures~\ref{flux}--\ref{fig:geoeff}.  After integrating over the
neutrino energy in \Eref{eq:stau flux: simple}, we obtain
approximate stau fluxes at the detector, summarized in
\Tref{tab:approx} for the ranges of incident neutrino fluxes given 
in \Sref{sec:nuflux}.

\begin{table}
\begin{center}
\vspace*{0.4cm}
\begin{tabular}{ccccc}\hline
Model & WB bound & WB GRB & Atm. Prompt & Atm. Conv.\\
\hline
{\bf I} & $<3.2$ & 0.20 & 0.012--0.73 & 0.0023--0.0038\\
{\bf II} & $<1.2$ & 0.066 & 0.0028--0.16 & 0.00028--0.00048\\
{\bf SPS7} & $<0.88$ & 0.045 & 0.0018--0.099 & 0.00014--0.00024\\
\boldmath{$\varepsilon$} & $<0.71$ & 0.034 & 0.0011--0.062 &
0.000069--0.00012\\
\hline
\end{tabular}
\end{center}
\caption{Stau fluxes from various neutrino sources for the four
benchmark supersymmetry models, {\it in the thick target
approximation}, in units of km$^{-2}$ yr$^{-1}$.}
\label{tab:approx}
\end{table}

So far we worked with the assumption that $y = 0.5$; i.e.,
each stau carries half of the incident neutrino energy.
However, especially when including the complex pattern of chain decay of
squarks and
sleptons into staus, a smaller fraction of the maximally available energy
is expected to be carried by the staus, implying a larger value for
$y$. The precise value for $y$ is model-dependent, and its detailed
evaluation is beyond the scope of the present analysis. We thus
investigate the dependence of the stau flux on the $y$ parameter, for
simplicity under the thick target approximation.
Our calculations show that the flux of staus, with an incoming flux
saturating the WB bound, and with our benchmark supersymmetric model
{\bf I}, is 3.2, 2.7, and 1.5 km$^{-2}$
yr$^{-1}$ for $y = 0.5$, 0.7, and 0.95, respectively.
This shows that the stau flux changes at most a factor of
$\sim$2 for a wide range of $y$, which is well within other model
uncertainties.
The weak dependence we find stems from the fact that a larger value of
$y$ requires
larger neutrino energies to produce staus with a certain energy. In
turn, at larger energies the incident flux is
smaller, but the cross section ratio is larger.
The cancellation of these two effects results in the mild dependence we
find.
The same argument also applies when we evaluate the flux more
accurately in the next subsection, where we thus again use the
assumption $y = 0.5$, for simplicity.

\subsection{Results of numerical integration: stau spectrum}\label{sub:integra}

We now solve \Eref{eq:detection flux} numerically to obtain a
more precise spectrum and rate estimate of staus at the detector, as
well as to check that the approximation made in the previous
subsection is reasonable.  Before giving the final flux estimates, we
start by investigating the generic structure of the integrand of
\Eref{eq:detection flux}.

\begin{figure}
\begin{center}
\includegraphics[width=15cm]{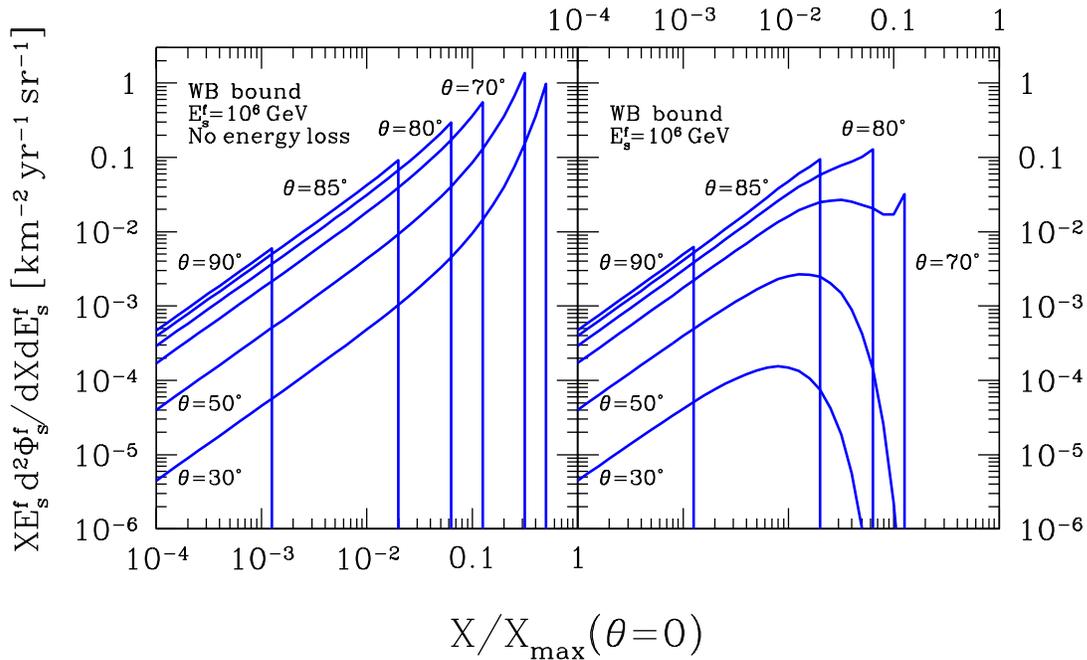}
\vspace*{-6mm}
\caption{The contribution to the stau events from a given column depth
  $X$, at an energy $E_s^f = 10^6$ GeV, for an incident neutrino flux
  saturating the WB bound. Each curve corresponds to a different value
  of the nadir angle $\theta$ as labeled. Left panel is shown with the
  assumption that {\it there is no energy loss for staus}, just for
  illustration purpose. In the right panel, proper energy losses are
  included.}
\label{fig:xdist}
\vspace{-6mm}
\end{center}
\end{figure}

\Fref{fig:xdist} shows the integrand of \Eref{eq:detection
flux} as a function of the column depth $X$ for various values of the
nadir angle $\theta$, assuming $E_s^f = 10^6$ GeV.
For definiteness, we show here the result only for a neutrino
injection model saturating the WB upper limit.
From the right panel of this figure, one can see that the stau events
are dominated by
directions with large enough nadir angle,
specifically with $\theta \gtrsim 70^\circ$,
so that the staus can reach the detector.
On the other hand, for very large nadir angles, the contributions
to the total event rate are modest, because Earth is not completely
opaque in those directions, even at these high energies.

\begin{figure}[!ht]
\begin{center}
\includegraphics[width=14cm]{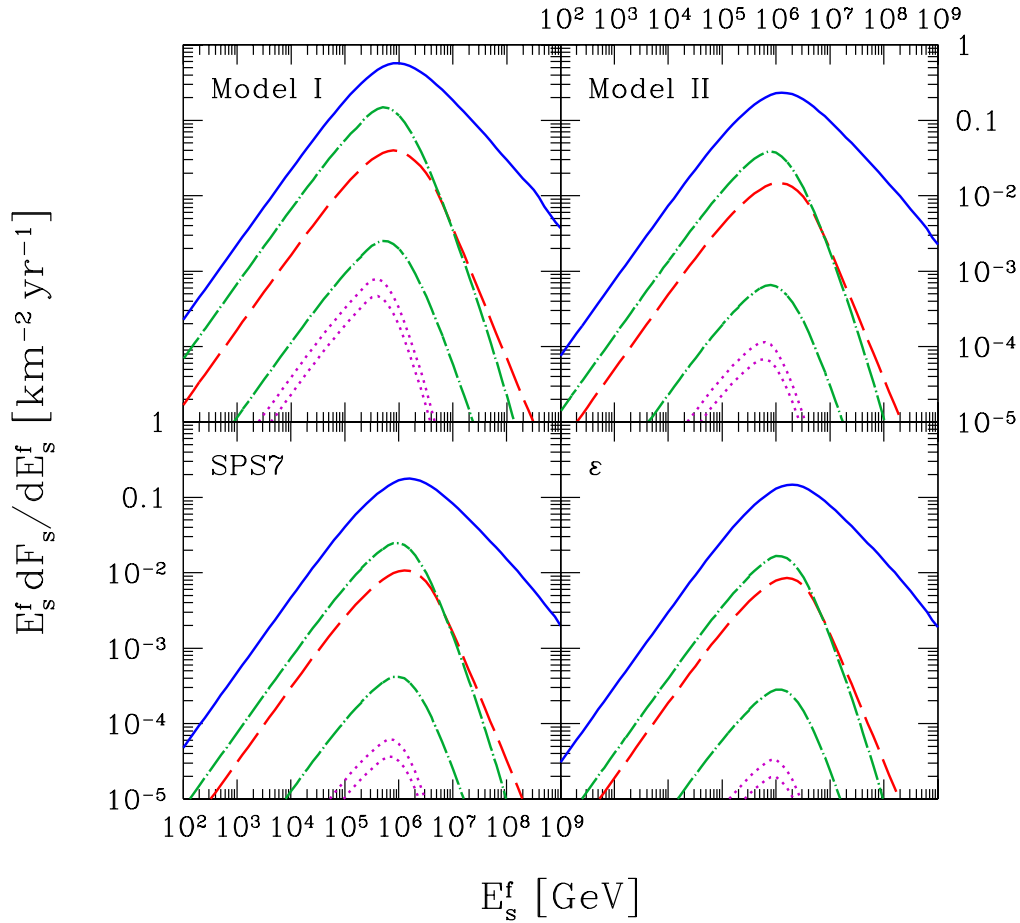}
\vspace*{-5mm}
\caption{The differential stau flux at the detector for the four benchmark
  supersymmetry models. Each line corresponds to a different incident
  neutrino flux model: WB upper limit (blue solid), WB GRB (red
  dashed), atmospheric prompt (green dot-dashed), and atmospheric
  conventional (magenta dotted).}
\label{fig:det_flux}
\vspace{-5mm}
\end{center}
\end{figure}

\Fref{fig:det_flux} shows the differential stau fluxes as a function of 
final stau energy.  First, we note that nearly all of the detectable
staus are well above the threshold required to be relativistic, which
is a stau energy comparable to the stau mass.  Only relativistic
charged particles produce the \v{C}erenkov light that IceCube can
measure.

The energy loss of relativistic particles may be dominated by
ionization or radiation, depending on whether the $\alpha$ term or the
$\beta$ term dominates in \Eref{eq:dEdX}, respectively.  This
transition for muons occurs at an energy $\alpha/\beta\simeq 500$~GeV.
For staus, it occurs at an energy a factor $\simeq m_s/m_\mu$ higher,
i.e., at least $10^6$~GeV.  The energy loss associated with
\v{C}erenkov radiation is always negligible; on the other hand, the
\v{C}erenkov radiation per unit length is the same for all
relativistic particles.  Thus, for particles at {\it any} energy in
the relativistic ionization-dominated regime, all tracks will look the
same in IceCube.

At higher energies, in the radiation dominated regime, there is
additional \v{C}erenkov radiation arising from relativistic electrons
and positrons created in hard radiative processes.  In this regime,
one can indeed tell the energy of the primary particle by the
intensity of the total \v{C}erenkov radiation.  For most of the
relevant final stau energies shown in \Fref{flux}, the staus will at
most be only slightly in the radiative regime, and so all stau tracks
going through IceCube will be indistinguishable from each other (and
from low-energy muons).  
While the total energy deposited in the detector is much smaller
for staus than it is for low-energy muons, this is irrelevant for
IceCube, which detects only the \v{C}erenkov light.

We argue that low-energy but relativistic stau pairs could also be detected
(albeit without energy measurement), giving a sizable event rate.
Recall that while it might be difficult to distinguish between staus
and muons on the basis of a single-particle detection, it would still
be possible if we use dual-track events: since staus propagate over
much longer distances in Earth than muons, tracks entering the
detector simultaneously are expected to be
well-separated~\cite{Albuquerque:2003mi,Ahlers:2006pf,Albuquerque:2006am}.
The careful analysis of Ref.~\cite{Albuquerque:2006am} shows that the
separation distribution of stau pairs ranges widely from 50 m to 1 km,
and that it peaks around $\sim$500 m.
On the other hand, the separation distribution for di-muons---the main
background for the stau pair track---peaks at around 10 m, and  
essentially no di-muon events with $> 50$ m separation are expected.
Therefore, this criterion rejects almost all background di-muon
events, but would capture a large fraction of stau events (typically
$>50\%$).  Thus, we are interested in the stau flux integrated over
energies larger than the relativistic threshold---which is very small,
$E_{s,{\rm th}} \simeq 300$ GeV---and where our results are least
sensitive.  This is clear from \Fref{fig:det_flux}.

\Tref{tab:true} shows the expected stau flux at the detector,
obtained by integrating the spectrum above 300 GeV.  Comparing with
the results of \Tref{tab:approx}, where we used the thick target
approximation, we find the two results are consistent with each other
within a factor of 2, justifying the reliability of the thick target
approximation.

\begin{table}[!ht]
\begin{center}
\begin{tabular}{ccccc}\hline
Model & WB bound & WB GRB & Atm. Prompt & Atm. Conv.\\
\hline
{\bf I} & $<2.2$ & 0.13 & 0.0070--0.41 & 0.0010--0.0018\\
{\bf II} & $<0.93$ & 0.049 & 0.0019--0.11 & 0.00017--0.00029\\
{\bf SPS7} & $<0.70$ & 0.035 & 0.0012--0.070 & 0.000091--0.00015\\
\boldmath{$\varepsilon$} & $<0.57$ & 0.027 & 0.00080--0.047 &
0.000049--0.000082\\
\hline
\end{tabular}
\end{center}
\caption{The stau flux from various neutrino sources for the four
  supersymmetry benchmark models, obtained from the {\it exact
  numerical integration of \Eref{eq:detection flux}},
  in units of km$^{-2}$ yr$^{-1}$.}
\label{tab:true}
\end{table}

Note that the flux of conventional atmospheric neutrinos is not
totally isotropic, but peaks in the horizontal direction by almost one
order of magnitude compared to other directions~\cite{Candia:2003ay}.
(The prompt flux is isotropic.)  This is an important effect because
most of the staus reaching the detector arose from neutrinos from
horizontal directions (\Fref{fig:xdist}).  Therefore, our results in
Tables~\ref{tab:approx} and~\ref{tab:true} for atmospheric neutrinos
would be larger by a factor of $\sim$3, as these results were obtained
with a direction-averaged incident neutrino flux.  

As a consequence, given that the predicted stau flux could be as large
as $\sim$1 km$^{-2}$ yr$^{-1}$ and that one expects essentially no
background from muon pair events, the search for stau pair tracks is warranted
in the actual data.
Our discussion here shows that a significant fraction of stau events
could possibly come from atmospheric prompt-decay neutrinos
(if the flux is close to the current upper bound), regardless of the
assumed supersymmetry models.

\subsection{Stau flux from nucleon--nucleon collisions}\label{sec:NN}

The stau pair flux $\Phi_{s}$ produced from a differential flux of
primary high-energy nucleons\footnote{Secondary nucleons and other
hadrons contribute a small fraction of the incoming flux, and taking
them into account does not affect our conclusions.} ${\rm
d}\Phi_N/{\rm d}E_N$ colliding with atmospheric nuclei is given by
\begin{equation}
\frac{{\rm d}\Phi_s}{{\rm d}E_s}\simeq 
\frac{\sigma_{NN,{\rm SUSY}}}{\sigma_{NN,{\rm tot}}}
\frac{{\rm d}\Phi_N}{{\rm d}E_N}
\frac{{\rm d}E_N}{{\rm d}E_s},
\end{equation}
because the thick target approximation (introduced in
\Sref{sec:thicktarget}) is very good for $NN$ interaction.
The symbols $\sigma_{NN,{\rm tot}}$ and $\sigma_{NN, {\rm SUSY}}$
indicate the total nucleon--nucleon interaction cross section and the
cross section into {\em any supersymmetric particle pair},
respectively.  To reiterate, since direct decays into gravitinos are
strongly suppressed by gravitational couplings, all final state $R$
parity odd particles decay into the NLSP, {\em i.e.}, lightest stau
pairs. For the total $NN$ cross section, we assume the parameterization for the hadron-air total cross section
\cite{Illana:2006xg}
\begin{equation}
A\frac{\sigma_{NN, {\rm tot}}}{\rm mb}\ \approx\
185+13.3\log\left(\frac{E_N}{\rm
  GeV}\right)+0.08\log^2\left(\frac{E_N}{\rm GeV}\right),
\end{equation}
where $A\simeq14.6$ is the average number of nucleons in a nucleus of air. We approximate the nucleon--nucleon cross section into supersymmetric
particles with the proton--proton cross section, and we compute the
latter using {\sc Prospino2.0}~\cite{Beenakker:1996ed}. For the
incoming nucleon flux we use the estimate in figure~1 of
Ref.~\cite{Illana:2006xg}.

Quark--anti-quark processes can produce stau pairs directly, unlike
neutrino--quark processes, where the final state has a larger
threshold as the final state must contain a typically-heavier squark.
Hence the kinematic threshold, as a function of the primary particle
energy, is lower in nucleon--nucleon collisions than in
neutrino--nucleon collisions.  Also, the subsequent occurrence of
various supersymmetric particle thresholds at larger and larger
masses, including particles featuring large degeneracy factors (such
as squarks), implies a more rapidly growing behavior for the
$\sigma_{NN,{\rm SUSY}}$ cross section than that for $\sigma_{\nu
N,{\rm SUSY}}$.  As a consequence, we find that for the models under
consideration here, $E_p\ {\rm d}N_s/{\rm d}E_p$ is almost constant
over several orders of magnitude in $E_p$.

The final flux of staus is, however, dramatically suppressed by the
ratio $\sigma_{NN,{\rm SUSY}}/\sigma_{NN, {\rm
tot}}$~\cite{Byrne:2002ri}, even taking into account multiplicity
effects in stau pair production or proton re-interactions.  In
particular, for the models we consider here, where the strongly
interacting supersymmetric particles are typically much more massive
than the NLSP, the combination of threshold effects and of the rapidly
decreasing flux of incident primary protons leads to dramatically less
optimistic predictions than those recently reported in
Ref.~\cite{Ahlers:2007js}.  There, the authors considered squarks and
gluinos with extremely low masses (150 and 300~GeV), while the
theoretically-motivated benchmark models we use here feature squark
and gluino masses between 600~GeV and 1~TeV.  While we agree with the numerical results reported in Ref.~\cite{Ahlers:2007js} when making the same assumptions on squark and gluino masses, we obtain much lower figures for the benchmark models we adopt here. Namely, we find
that for the two most optimistic models, {\bf I} and {\bf II}, we
predict a stau flux in IceCube from nucleon--nucleon interactions of
$10^{-4}$ and $3\times10^{-6}$ per km$^{2}$ per year,
respectively---much lower than even the contribution from conventional
atmospheric neutrinos.  We believe that this relative smallness compared
with that from the atmospheric incident neutrinos would be a rather
model-independent feature.


\section{Role of the supersymmetric particle spectrum}\label{sec:susy}

In the previous sections we focused on specific supersymmetric models.
We now wish to address the model-independent question of how the stau
pair rate at neutrino telescopes depends upon the supersymmetric
particle masses.  As we already pointed out, the stau pair flux
depends upon the $\nu N\rightarrow {\rm SUSY}$ cross section: the
relevant masses entering the cross section are those of the heavy pair
produced, and those of the supersymmetric partners of the electroweak
gauge bosons which mediate the charged- and neutral-current
interactions responsible for slepton--squark production. In addition,
the number of produced stau pairs as a function of incoming
neutrino energy crucially depends on the final state kinematic
threshold: the larger the latter, the smaller the flux of incoming
neutrinos that can lead to stau pair production, hence a smaller
expected stau pair rate.

\begin{figure}[t]
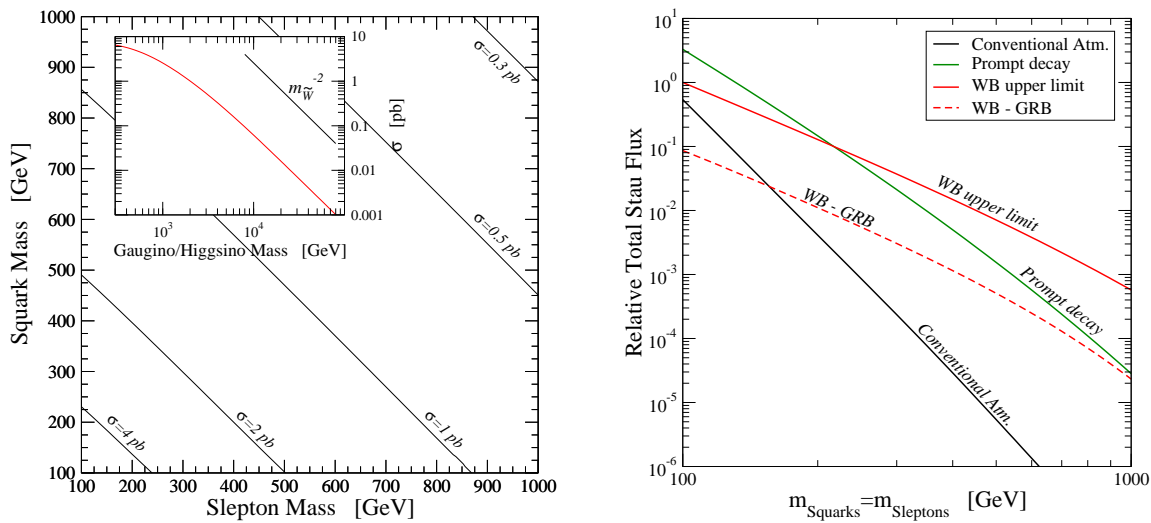

\begin{center}
\mbox{\includegraphics[width=7.3cm,clip]{figures/iso.eps}\qquad
      \includegraphics[width=7cm,clip]{figures/flux_vs_susy.eps}}
\caption{{\em Left:} Isovalue curves (labeled) of the neutrino--proton 
cross section for supersymmetric particle pair production at incident
$E_\nu=10^8$ GeV, in the plane defined by the slepton and squark
masses (for simplicity we assume degenerate sleptons and degenerate
squarks; see text for details of the supersymmetric models).  The
inset illustrates the behavior of the cross section, again at $E_\nu =
10^8$ GeV, as a function of a common gaugino/higgsino mass
($m_1=m_2=\mu$). {\em Right:} Contributions from conventional, prompt
decay and extra-galactic high-energy neutrinos to the total stau flux,
as a function of the supersymmetric scalars' (degenerate) masses),
normalized to be relative.}
\label{susy}
\end{center}
\end{figure}

To explore quantitatively the statements above, we employ a convenient
phenomenological parameterization of the supersymmetric setup at the
low-energy scale, and no longer rely on a specific supersymmetry
breaking framework.  The left panel of \Fref{susy} shows the stau
production cross section variation in high-energy neutrino--proton
collisions in the plane defined by the slepton (x-axis) and squark
(y-axis) masses.  For definiteness, we fix the relevant gaugino
(respectively, bino and wino) masses to $m_1 = 1$ TeV and $m_2 = 2$
TeV.  We choose $\mu = 1$ TeV, $m_A = 500$ GeV, $\tan\beta = 10$, set
all trilinear scalar couplings to zero, and further assume that the
soft supersymmetry breaking scalar masses of sleptons and (separately)
of squarks are degenerate, CP-conserving and flavor-diagonal.  In the
figure, we show iso-level curves at fixed values of the production
cross section for staus at an incident neutrino energy of $10^8$ GeV,
in order to avoid production threshold effects.  As the figure
illustrates, cross section variation is very mild well above
threshold.  Quantitatively, the effect varies within little more than
one order of magnitude for scalar masses varied between 100 and 1000
GeV.

In the inset, we show how the cross section scales the masses of
$t$-channel supersymmetric particles (neutralinos and charginos)
exchanged in squark--slepton pair production from neutrino--proton
collisions.  Neutralino and chargino masses are entirely determined at
tree level by the gaugino soft supersymmetry breaking masses $m_1$ and
$m_2$ and the higgsino mass term $\mu$, and by $\tan\beta$.  For
simplicity we assume a common `{\em gaugino/higgsino}' mass scale,
$m_{\wt W}$, defined as the common value of $m_1 = m_2 = \mu$.  We
employ a common scalar mass $m_S$ for both sleptons and squarks of 300
GeV, and set all other low-scale supersymmetric parameters as in the
rest of the figure ($m_A=500$ GeV, $\tan\beta=10$, all trilinear
scalar couplings zero).  As we illustrate in the inset, beyond the
scalar mass scale ($m_{\wt W}\lesssim m_S$) where kinematic effects
play a non-trivial role, cross section scaling goes like the
gaugino/higgsino mass scale to the power $-2$.  This can be
analytically understood, since
\begin{eqnarray}
\nonumber \sigma_{\nu p}&\sim&\int_{4m_S^2/s}^1{\rm d}x\int_0^{xs}\
{\rm d}Q^2\ \frac{{\rm d}^2\sigma_{\nu p}}{{\rm d}x\ {\rm
    d}Q^2}\sim\int_{4m_S^2/s}^1{\rm d}x\int_0^{xs}\ {\rm
  d}Q^2\ \frac{[x\cdot q(x,Q^2)]}{\left(Q^2+m_{\wt
    W}^2\right)^2}\\
\label{eq:final}&=&\int_{4m_S^2/s}^1\frac{x^{-1/3}\ s\ {\rm d}x}{\left(m_{\wt W}^4+xsm_{\wt W}^2\right)^2}.
\end{eqnarray}
where, in \Eref{eq:final}, we made use of the fact that in the large
$E_\nu$ regime, $[x\cdot q(x,Q^2)]\sim x^{-1/3}$~\cite{Gandhi:1998ri}.
One thus gets:
\begin{equation}
\sigma_{\nu p}\sim\frac{1}{m_{\wt W}^2}\left(\frac{s}{4m_S^2}\right)^{1/3},
\end{equation}
which explains both the scaling in the inset of \Fref{susy} and of
$\sigma_{\nu p}$ in \Fref{xsec}.

As the supersymmetric particle spectrum gets heavier, not only does
the neutrino--proton cross section become smaller (left panel of
\Fref{susy}), but more importantly the shift in incident neutrino 
energy threshold for stau production strongly suppresses the final
stau flux. This depends on the dramatic dependence of the flux of
incident neutrinos on energy, as illustrated in our \Fref{flux}.  The
right panel of \Fref{susy} quantifies this trend.  There we show the
relative contribution to the total stau flux from various incident
neutrino fluxes as a function of the common squark and slepton masses
$m_S$.  The fluxes are normalized to that resulting from the incident
WB upper limit of extra-galactic neutrino flux and $m_S=100$ GeV.  We
set all the supersymmetric parameters to the same values as in the
left panel.
Comparing the relative flux for various origins summarized in
\Tref{tab:true} with that shown in the right panel of \Fref{susy}, we
can roughly estimate that the four benchmark models correspond to $m_S
\approx 500$ GeV in the context of this phenomenological approach.
In addition, recalling that the benchmark models predict stau flux of
$\sim$1 km$^{-2}$ yr$^{-1}$ for WB neutrino bound, this suggests that
one can expect a stau flux well above 1 km$^{-2}$ yr$^{-1}$, provided 
slepton and squark masses are smaller than 500 GeV.

The different neutrino flux scaling with energy dictates that the
relative importance of the various neutrino sources depends on the
mass scale of the particles produced in the neutrino--proton
collision.  In particular, while with a very light spectrum the
contribution from conventional atmospheric neutrinos can be comparable
to (or even larger than) the extra-galactic neutrino component, for
$m_S\gtrsim0.5$~TeV the contribution from atmospheric neutrinos
becomes negligible.  Note that in that mass range the supersymmetric
particles are so heavy that the overall stau flux is extremely
suppressed, and likely undetectable.  On the other hand, the figure
illustrates that, in principle, prompt decay neutrinos can be the
dominant source of staus for almost any value of $m_S$ if the
extra-galactic neutrino flux is close to the GRB-derived range (dashed
red line in the figure) rather than the WB upper limit.  In addition,
even conventional neutrinos contribute a stau flux of the same order
of magnitude as that expected from the WB upper limit on astrophysical
neutrinos, as long as the supersymmetric scalars mass scale is below
200~GeV. If both the prompt-decay neutrino flux and the extra-galactic
neutrino flux are maximal, prompt neutrinos contribute at the same
level as extra-galactic neutrinos for $m_S\lesssim 400$~GeV, and
dominate for $m_S$ below $200$~GeV.  Since a detectable signal is
expected only for a light supersymmetric spectrum, this leads us to
the following prediction: if the signal discussed here is indeed
detected, a very sizable fraction of it will originate from
conventional and prompt-decay neutrinos.


\section{Conclusions}\label{sec:concl}

We reassessed the flux of meta-stable staus produced by
neutrino--nucleon and nucleon--nucleon interactions that might be
detectable at km$^3$ neutrino telescopes.  We derived the flux of
staus from first principles, and showed that, under the approximation
that Earth is opaque to very high energy neutrinos, the number of
staus at the detector is given by a simple integral over the neutrino
energy of the product of three factors: the incident neutrino flux,
the ratio of the neutrino--nucleon cross section into supersymmetric
particle pairs over the total neutrino--nucleon cross section, and a
geometric efficiency factor.  We showed that this approximation
reproduces an exact numerical computation within a factor 2, which in
turn is much better than the level of our knowledge of the first two
factors entering the stau flux computation---namely, the incident
neutrino flux and the features of the supersymmetric particle setup.

We focused on each of the factors relevant to the computation of the
final flux of staus.  We concentrated on four well-motivated
supersymmetric benchmark models, and independently evaluated the
relevant production cross sections.  We pointed out that
previously-neglected atmospheric neutrinos from prompt charmed meson
decays could give a potentially large stau flux, even in the absence
of a (yet to be discovered) astrophysical high-energy neutrino flux.
This will depend on the prompt atmospheric neutrino flux being at the
upper end of the theoretically expected range.
Nucleon--nucleon processes, even for the most optimistic benchmark
models, would not contribute sizably to the final stau flux.  Finally,
we numerically and analytically studied how the relevant cross
sections depend on the supersymmetric model mass spectrum, and how the
relative importance of primary neutrino sources depends on the mass
scale of supersymmetric scalars.  In particular, we predict that if
the signal discussed here is indeed detected, a very sizable fraction
of it would originate from conventional and prompt-decay neutrinos.


\section*{Acknowledgments}

This work was supported by the Sherman Fairchild Foundation (SA); CCAPP,
The Ohio
State University, and NSF CAREER grant PHY-0547102 (JFB); DoE grants
DE-FG03-92-ER40701, DE-FG02-05ER41361, and NASA grant NNG05GF69G (SP); and
DoE grant DE-FG02-91ER40685 (DR).  We would like to thank Francis
Halzen, Chris Quigg, and Xerxes Tata for enlightening discussions and Markus Ahlers for useful correspondence.




\section*{References}

\begin{thebibliography}{200}

\bibitem{Fukugita:2006rm}
  M.~Fukugita, K.~Ichikawa, M.~Kawasaki and O.~Lahav,
  Phys.\ Rev.\ D {\bf 74}, 027302 (2006).

\bibitem{Spergel:2006hy}
  D.~N.~Spergel {\it et al.} [WMAP Collaboration],
  Astrophys.\ J.\ Suppl.\  {\bf 170}, 377 (2007).

\bibitem{Bergstrom:2000pn}
  L.~Bergstrom,
  Rept.\ Prog.\ Phys.\  {\bf 63}, 793 (2000); \\
  G.~Bertone, D.~Hooper and J.~Silk,
  Phys.\ Rept.\  {\bf 405}, 279 (2005).

\bibitem{Jungman:1995df}
  G.~Jungman, M.~Kamionkowski and K.~Griest,
  Phys.\ Rept.\  {\bf 267}, 195 (1996).

\bibitem{Cheng:2002ej}
  H.~C.~Cheng, J.~L.~Feng and K.~T.~Matchev,
  Phys.\ Rev.\ Lett.\  {\bf 89}, 211301 (2002).

\bibitem{Hooper:2007qk}
  D.~Hooper and S.~Profumo,
  Phys.\ Rept.\  {\bf 453}, 29 (2007).

\bibitem{Wolfram:1978gp}
  S.~Wolfram,
  Phys.\ Lett.\ B {\bf 82}, 65 (1979).

\bibitem{Munoz:2003gx}
  C.~Munoz,
  Int.\ J.\ Mod.\ Phys.\ A {\bf 19}, 3093 (2004).

\bibitem{Bergstrom:1998hd}
  L.~Bergstrom,
  New Astron.\ Rev.\  {\bf 42}, 245 (1998).

\bibitem{Drees:2000he}
  M.~Drees, Y.~G.~Kim, M.~M.~Nojiri, D.~Toya, K.~Hasuko and T.~Kobayashi,
  Phys.\ Rev.\  D {\bf 63}, 035008 (2001).

\bibitem{Baer:2003wx}
  H.~Baer, C.~Balazs, A.~Belyaev, T.~Krupovnickas and X.~Tata,
  JHEP {\bf 0306}, 054 (2003).

\bibitem{Weiglein:2004hn}
  G.~Weiglein {\it et al.}  [LHC/LC Study Group],
  Phys.\ Rept.\  {\bf 426}, 47 (2006).

\bibitem{Masiero:2004ft}
  A.~Masiero, S.~Profumo and P.~Ullio,
  Nucl.\ Phys.\  B {\bf 712}, 86 (2005).

\bibitem{Baltz:2006fm}
  E.~A.~Baltz, M.~Battaglia, M.~E.~Peskin and T.~Wizansky,
  Phys.\ Rev.\ D {\bf 74}, 103521 (2006).

\bibitem{howie}
  H.~Baer and X.~Tata,
  ``Weak Scale Supersymmetry: From Superfields to Scattering Events'',
  Cambridge University Press, 2006.

\bibitem{Blumenthal:1982mv}
  G.~R.~Blumenthal, H.~Pagels and J.~R.~Primack,
  Nature {\bf 299}, 37 (1982).

\bibitem{Feng:2003nr}
  J.~L.~Feng, A.~Rajaraman and F.~Takayama,
  Phys.\ Rev.\ D {\bf 68}, 085018 (2003).

\bibitem{Feng:2003xh}
  J.~L.~Feng, A.~Rajaraman and F.~Takayama,
  Phys.\ Rev.\ Lett.\  {\bf 91}, 011302 (2003).

\bibitem{lep2}
  F.~Cerutti et al. [LEP2 SUSY Working Group], LEPSUSYWG/02-05.1, 
{\tt http://lepsusy.web.cern.ch/lepsusy/www/stable-summer02/stable-208.html}

\bibitem{tevatron}
  D.~Acosta {\it et al.}  [CDF Collaboration],
  Phys.\ Rev.\ Lett.\  {\bf 90}, 131801 (2003).

\bibitem{Fairbairn:2006gg}
  M.~Fairbairn \etal, 
  Phys.\ Rept.\  {\bf 438}, 1 (2007).

\bibitem{chargeddecay} See {\em e.g.}
  X.~L.~Chen and M.~Kamionkowski,
  Phys.\ Rev.\  D {\bf 70}, 043502 (2004).

\bibitem{Zhang:2007zzh}
  L.~Zhang, X.~Chen, M.~Kamionkowski, Z.~g.~Si and Z.~Zheng,
  Phys.\ Rev.\  D {\bf 76}, 061301 (2007)

\bibitem{Cembranos:2007fj}
  J.~A.~R.~Cembranos, J.~L.~Feng and L.~E.~Strigari,
  arXiv:0704.1658 [astro-ph].

\bibitem{Pierpaoli:2003rz}
  E.~Pierpaoli,
  Phys.\ Rev.\ Lett.\  {\bf 92}, 031301 (2004).

\bibitem{Sigurdson:2003vy}
  K.~Sigurdson and M.~Kamionkowski,
  Phys.\ Rev.\ Lett.\  {\bf 92}, 171302 (2004).

\bibitem{Profumo:2004qt}
  S.~Profumo, K.~Sigurdson, P.~Ullio and M.~Kamionkowski,
  Phys.\ Rev.\  D {\bf 71}, 023518 (2005).

\bibitem{BBNLDP} See,{\em  e.g.}, 
  K.~Kohri and F.~Takayama,
  Phys.\ Rev.\  D {\bf 76}, 063507 (2007);
  M.~Kawasaki, K.~Kohri and T.~Moroi,
  Phys.\ Lett.\  B {\bf 649}, 436 (2007);
  T.~Jittoh, K.~Kohri, M.~Koike, J.~Sato, T.~Shimomura and M.~Yamanaka,
  arXiv:0704.2914 [hep-ph].

\bibitem{Jedamzik:2007cp}
  K.~Jedamzik,
  arXiv:0707.2070 [astro-ph].

\bibitem{BBNLDP2}
  D.~Cumberbatch, K.~Ichikawa, M.~Kawasaki, K.~Kohri, J.~Silk and G.~D.~Starkman,
  arXiv:0708.0095 [astro-ph];
  and references therein.

\bibitem{Feng:2004yi}
  J.~L.~Feng and B.~T.~Smith,
  Phys.\ Rev.\ D {\bf 71}, 015004 (2005)
  [Erratum-ibid.\ {\bf 71}, 0109904 (2005)].

\bibitem{Albuquerque:2003mi}
  I.~Albuquerque, G.~Burdman and Z.~Chacko,
  Phys.\ Rev.\ Lett.\  {\bf 92}, 221802 (2004).

\bibitem{Ahlers:2006pf}
  M.~Ahlers, J.~Kersten and A.~Ringwald,
  JCAP {\bf 0607}, 005 (2006).

\bibitem{Albuquerque:2006am}
  I.~F.~M.~Albuquerque, G.~Burdman and Z.~Chacko,
  Phys.\ Rev.\  D {\bf 75}, 035006 (2007).

\bibitem{Reno:2005si}
  M.~H.~Reno, I.~Sarcevic and S.~Su,
  Astropart.\ Phys.\  {\bf 24}, 107 (2005).

\bibitem{Huang:2006ie}
  Y.~Huang, M.~H.~Reno, I.~Sarcevic and J.~Uscinski,
  Phys.\ Rev.\  D {\bf 74}, 115009 (2006).

\bibitem{Ahlers:2007js}
  M.~Ahlers, J.~I.~Illana, M.~Masip and D.~Meloni,
  JCAP {\bf 0708}, 008 (2007).

\bibitem{Reno:2007kz}
  M.~H.~Reno, I.~Sarcevic and J.~Uscinski,
  arXiv:0710.4954 [hep-ph].

\bibitem{WBBound}
  E.~Waxman and J.~N.~Bahcall,
  Phys.\ Rev.\ D {\bf 59}, 023002 (1999).

\bibitem{AtmExp}
  K.~Daum {\it et al.}  [Frejus Collaboration.],
  Z.\ Phys.\ C {\bf 66}, 417 (1995); \\
  J.~Ahrens {\it et al.}  [AMANDA Collaboration],
  Phys.\ Rev.\ D {\bf 66}, 012005 (2002); \\
  Y.~Ashie {\it et al.}  [Super-Kamiokande Collaboration],
  Phys.\ Rev.\ D {\bf 71}, 112005 (2005).

\bibitem{WBGRB}
  E.~Waxman and J.~N.~Bahcall,
  Phys.\ Rev.\ Lett.\  {\bf 78}, 2292 (1997).

\bibitem{ABJet}
  S.~Ando and J.~F.~Beacom,
  Phys.\ Rev.\ Lett.\  {\bf 95}, 061103 (2005).

\bibitem{starburst}
  A.~Loeb and E.~Waxman,
  JCAP {\bf 0605}, 003 (2006); \\
  F.~W.~Stecker,
  Astropart.\ Phys.\  {\bf 26}, 398 (2007); \\
  T.~A.~Thompson, E.~Quataert, E.~Waxman and A.~Loeb,
  astro-ph/0608699.

\bibitem{AtmTheory}
  G.~D.~Barr, T.~K.~Gaisser, P.~Lipari, S.~Robbins and T.~Stanev,
  Phys.\ Rev.\ D {\bf 70}, 023006 (2004); \\
  G.~D.~Barr, T.~K.~Gaisser, S.~Robbins and T.~Stanev,
  Phys.\ Rev.\ D {\bf 74}, 094009 (2006); \\
  M.~Honda, T.~Kajita, K.~Kasahara and S.~Midorikawa,
  Phys.\ Rev.\ D {\bf 70}, 043008 (2004); \\
  M.~Honda, T.~Kajita, K.~Kasahara, S.~Midorikawa and T.~Sanuki,
  Phys.\ Rev.\  D {\bf 75}, 043006 (2007).

\bibitem{Candia:2003ay}
  J.~Candia and E.~Roulet,
  JCAP {\bf 0309}, 005 (2003).

\bibitem{Zas:1992ci}
  E.~Zas, F.~Halzen and R.~A.~Vazquez,
  Astropart.\ Phys.\  {\bf 1}, 297 (1993).

\bibitem{Pasquali:1998ji}
  L.~Pasquali, M.~H.~Reno and I.~Sarcevic,
  Phys.\ Rev.\  D {\bf 59}, 034020 (1999).

\bibitem{Gelmini:1999xq}
  G.~Gelmini, P.~Gondolo and G.~Varieschi,
  Phys.\ Rev.\  D {\bf 61}, 056011 (2000).

\bibitem{Costa:2000jw}
  C.~G.~S.~Costa,
  Astropart.\ Phys.\  {\bf 16}, 193 (2001).

\bibitem{Volkova:2001th}
  L.~V.~Volkova and G.~T.~Zatsepin,
  Phys.\ Atom.\ Nucl.\  {\bf 64}, 266 (2001)
  [Yad.\ Fiz.\  {\bf 64}, 313 (2001)].

\bibitem{Fiorentini:2001wa}
  G.~Fiorentini, V.~A.~Naumov and F.~L.~Villante,
  Phys.\ Lett.\  B {\bf 510}, 173 (2001).

\bibitem{Beacom:2004jb}
  J.~F.~Beacom and J.~Candia,
  JCAP {\bf 0411}, 009 (2004).

\bibitem{Martin:2003us}
  A.~D.~Martin, M.~G.~Ryskin and A.~M.~Stasto,
  Acta Phys.\ Polon.\ B {\bf 34}, 3273 (2003).

\bibitem{Achterberg:2006pw}
  A.~Achterberg  [IceCube Collaboration],
  astro-ph/0611597.

\bibitem{Feng:2004mt}
  J.~L.~Feng, S.~Su and F.~Takayama,
  Phys.\ Rev.\ D {\bf 70}, 075019 (2004).

\bibitem{sugra}
 J.~R.~Ellis, K.~A.~Olive, Y.~Santoso and V.~C.~Spanos,
  Phys.\ Lett.\  B {\bf 588}, 7 (2004); \\
  J.~R.~Ellis, A.~R.~Raklev and O.~K.~Oye,
  JHEP {\bf 0610}, 061 (2006); \\
  A.~Ibarra and S.~Roy,
  JHEP {\bf 0705}, 059 (2007).

\bibitem{gmsb}
  D.~A.~Dicus, B.~Dutta and S.~Nandi,
  Phys.\ Rev.\  D {\bf 56}, 5748 (1997); \\
  K.~Cheung, D.~A.~Dicus, B.~Dutta and S.~Nandi,
  Phys.\ Rev.\  D {\bf 58}, 015008 (1998); \\
  J.~L.~Feng and T.~Moroi,
  Phys.\ Rev.\  D {\bf 58}, 035001 (1998); \\
  P.~G.~Mercadante, J.~K.~Mizukoshi and H.~Yamamoto,
  Phys.\ Rev.\  D {\bf 64}, 015005 (2001).

\bibitem{Giudice:1998bp}
  G.~F.~Giudice and R.~Rattazzi,
  Phys.\ Rept.\  {\bf 322}, 419 (1999).

\bibitem{coan}
  S.~Ambrosanio and B.~Mele,
  Phys.\ Rev.\  D {\bf 55}, 1399 (1997)
  [Erratum-ibid.\  {\bf 56}, 3157 (1997)]; \\
  A.~V.~Gladyshev, D.~I.~Kazakov and M.~G.~Paucar,
  Mod.\ Phys.\ Lett.\  A {\bf 20}, 3085 (2005); \\
  T.~Jittoh, J.~Sato, T.~Shimomura and M.~Yamanaka,
  Phys.\ Rev.\  D {\bf 73}, 055009 (2006).

\bibitem{rhnlsp}
  T.~Asaka, K.~Ishiwata and T.~Moroi,
  Phys.\ Rev.\  D {\bf 73}, 051301 (2006); \\
  S.~K.~Gupta, B.~Mukhopadhyaya and S.~K.~Rai,
  Phys.\ Rev.\  D {\bf 75}, 075007 (2007).

\bibitem{Cembranos:2006gt}
  J.~A.~R.~Cembranos, J.~L.~Feng and L.~E.~Strigari,
  Phys.\ Rev.\  D {\bf 75}, 036004 (2007).

\bibitem{uedewpo}
  T.~Flacke, D.~Hooper and J.~March-Russell,
  Phys.\ Rev.\  D {\bf 73}, 095002 (2006)
  [Erratum-ibid.\  {\bf 74}, 019902 (2006)]; \\
  I.~Gogoladze and C.~Macesanu,
  Phys.\ Rev.\  D {\bf 74}, 093012 (2006).

\bibitem{Allanach:2002nj}
  B.~C.~Allanach {\it et al.},
in {\it Proc. of the APS/DPF/DPB Summer Study on the Future of Particle Physics (Snowmass 2001) } ed. N.~Graf,
{\it In the Proceedings of APS / DPF / DPB Summer Study on the Future of Particle Physics (Snowmass 2001), Snowmass, Colorado, 30 Jun - 21 Jul
2001, pp P125}
  [hep-ph/0202233].

\bibitem{DeRoeck:2005bw}
  A.~De Roeck, J.~R.~Ellis, F.~Gianotti, F.~Moortgat, K.~A.~Olive and L.~Pape,
  Eur.\ Phys.\ J.\  C {\bf 49}, 1041 (2007)
  [arXiv:hep-ph/0508198].

\bibitem{Ellis:2003dn}
  J.~R.~Ellis, K.~A.~Olive, Y.~Santoso and V.~C.~Spanos,
  Phys.\ Lett.\  B {\bf 588}, 7 (2004).

\bibitem{BBN}
  R.~H.~Cyburt, J.~R.~Ellis, B.~D.~Fields and K.~A.~Olive,
  Phys.\ Rev.\  D {\bf 67}, 103521 (2003); \\
  K.~Jedamzik,
  Phys.\ Rev.\ Lett.\  {\bf 84}, 3248 (2000);
  J.~L.~Feng, S.~f.~Su and F.~Takayama,
  Phys.\ Rev.\  D {\bf 70}, 063514 (2004);
  F.~D.~Steffen,
  JCAP {\bf 0609}, 001 (2006).

\bibitem{BBN2}
  M.~Kawasaki, K.~Kohri and T.~Moroi,
  Phys.\ Rev.\  D {\bf 71}, 083502 (2005);
  M.~Kawasaki, K.~Kohri and T.~Moroi,
  Phys.\ Lett.\  B {\bf 625}, 7 (2005).

\bibitem{CMB}
  W.~Hu and J.~Silk,
  Phys.\ Rev.\ Lett.\  {\bf 70}, 2661 (1993).

\bibitem{pospelov}
  M.~Pospelov,
  Phys.\ Rev.\ Lett.\  {\bf 98}, 231301 (2007).

\bibitem{Kaplinghat:2006qr}
  M.~Kaplinghat and A.~Rajaraman,
  Phys.\ Rev.\  D {\bf 74}, 103004 (2006)

\bibitem{jedamzik}
  K.~Jedamzik,
  arXiv:0707.2070 [astro-ph].

\bibitem{olive}
  R.~H.~Cyburt, J.~R.~Ellis, B.~D.~Fields, K.~A.~Olive and V.~C.~Spanos,
  JCAP {\bf 0611}, 014 (2006).

\bibitem{Cho:2006sx}
  G.~C.~Cho, K.~Hagiwara, J.~Kanzaki, T.~Plehn, D.~Rainwater and T.~Stelzer,
  Phys.\ Rev.\ D {\bf 73}, 054002 (2006).

\bibitem{Maltoni:2002qb}
  F.~Maltoni and T.~Stelzer,
  JHEP {\bf 0302}, 027 (2003).

\bibitem{SLHA}
  B.~Allanach, P.~Skands \etal,
  JHEP {\bf 0407}, 036 (2003); \\
  T.~Hahn,
  hep-ph/0408283; \\
  J.~A.~Aguilar-Saavedra {\it et al.},
  Eur.\ Phys.\ J.\  C {\bf 46}, 43 (2006).

\bibitem{Djouadi:2002ze}
  A.~Djouadi, J.~L.~Kneur and G.~Moultaka,
  hep-ph/0211331.

\bibitem{Pumplin:2002vw}
  J.~Pumplin, D.~R.~Stump, J.~Huston, H.~L.~Lai, P.~Nadolsky and W.~K.~Tung,
  JHEP {\bf 0207}, 012 (2002).

\bibitem{Carena:1998gd}
  M.~Carena, D.~Choudhury, S.~Lola and C.~Quigg,
  Phys.\ Rev.\ D {\bf 58}, 095003 (1998).

\bibitem{Gaisser:1985cm}
  T.~K.~Gaisser and T.~Stanev,
  Phys.\ Rev.\  D {\bf 31}, 2770 (1985).

\bibitem{Kistler:2006hp}
  M.~D.~Kistler and J.~F.~Beacom,
  Phys.\ Rev.\  D {\bf 74}, 063007 (2006).

\bibitem{Gandhi:1995tf}
  R.~Gandhi, C.~Quigg, M.~H.~Reno and I.~Sarcevic,
  Astropart.\ Phys.\  {\bf 5}, 81 (1996).

\bibitem{Illana:2006xg}
  J.~I.~Illana, M.~Masip and D.~Meloni,
  Phys.\ Rev.\  D {\bf 75}, 055002 (2007).

\bibitem{Beenakker:1996ed}
  W.~Beenakker, R.~Hopker and M.~Spira,
  hep-ph/9611232.

\bibitem{Byrne:2002ri}
  M.~Byrne, C.~F.~Kolda and P.~Regan,
  Phys.\ Rev.\  D {\bf 66}, 075007 (2002).

\bibitem{Gandhi:1998ri}
  R.~Gandhi, C.~Quigg, M.~H.~Reno and I.~Sarcevic,
  Phys.\ Rev.\ D {\bf 58}, 093009 (1998).

\end{thebibliography}
\end{document}